\pdfoutput=1

\documentclass[journal]{IEEEtran}
\ifCLASSINFOpdf
   \usepackage[pdftex]{graphicx}
   \graphicspath{{../pdf/}{../jpeg/}}
   \DeclareGraphicsExtensions{.pdf,.jpeg,.png}
\else
   \usepackage[dvips]{graphicx}
   \graphicspath{{../eps/}}
   \DeclareGraphicsExtensions{.eps}
\fi
\usepackage[cmex10]{amsmath}
\usepackage {amssymb, cases}
\newtheorem{theorem}{Theorem}

\newtheorem{proposition}{Proposition}

\usepackage{ algorithmic}
\usepackage{ algorithm}

\usepackage{array}

\ifCLASSOPTIONcompsoc
  \usepackage[caption=false,font=normalsize,labelfont=sf,textfont=sf]{subfig}
\else
  \usepackage[caption=false,font=footnotesize]{subfig}
\fi
\begin{document}

\title{A Closed-form Transceiver Design for Interference Alignment and Cancellation (IAC) in MIMO Interference Channel}
%
%
%

\author{Xin~Qu~and~Chung G.~Kang,~\IEEEmembership{Senior Member,~IEEE}}
\maketitle

\begin{abstract}
For a $K$-user interference channel, the degree of freedom (DoF) which can be achieved through interference alignment (IA) is constrained to signal space dimension governed by the number of Tx/Rx antennas. To overcome this problem, IA can be combined with interference cancellation (IC), involving a new receiver architecture associated with signaling over backhaul links among the different users, as another interference mitigation scheme which is referred to as interference alignment and cancellation (IAC). In our earlier work, by proposing an IAC graph, we have derived the necessary and sufficient conditions for the existence of closed-form solutions for IAC subject to the given DoF requirement for individual user. Furthermore, we have also shown that it can achieve the theoretically maximum possible DoF, which is $2M$ for MIMO system with $M$ Tx/Rx antennas. Following our previous works on IAC, we aim to investigate the design criteria to obtain such closed-form transceivers when they exist. We first develop a general closed-form IAC transceiver design for any given DoF requirement of individual user and then, we specify how the optimal IAC transceiver can be designed to achieve the theoretically maximum DoF of $2M$, beating the performance of conventional IA with much less computational complexity.
\end{abstract}


%
\IEEEpeerreviewmaketitle

\section{Introduction}
%
%
%
%
\IEEEPARstart{I}{n} a $K$-user multiple input multiple output (MIMO) Gaussian interference channel, a vector space interference alignment (IA) technique with neither time nor frequency diversity prevents interference on the desired user by causing the interference and user signals to use the different spaces provided by multiple Tx/Rx antennas. At the $k$-th Tx/Rx pair, a plurality of interference signals from all other $\left( K-1 \right)$ Tx/Rx pairs are aligned to reduce the occupied space, while the remaining interference-free space is preserved for desired signals. With a limited number of Tx/Rx antennas, the interference subspace dimension reduced by IA is constrained by two factors: the maximum number of interference signals transmitted from one of the interfering sources and the aligned level for interference from the remaining $\left( K-1 \right)$ interfering sources. First, the dimension of interference subspace after reduction would be no less than the value given by the first factor, since the interference signals from the same source could not be aligned. Then, the reduced dimension depends on the number of Tx/Rx antennas as it determines the capability of alignment operation. In order to further reduce the occupied interference subspace while expanding signal subspace under the limited number of Tx/Rx antennas, therefore, interference cancellation (IC) can be combined with vector space IA as another type of interference mitigation technique, which is referred to as interference alignment and cancellation (IAC) [1]. Under the help of IC, the number of interfering sources can be dramatically reduced in the receivers. Therefore, more dimension of signal subspace would be preserved by IAC, which would gain more degrees of freedom (DoFs) over IA for a $K$-user MIMO interference channel.

In the classical IA scheme, let ${{\mathbf{V}}_{j}}$ and ${{\mathbf{U}}_{k}}$ denote the precoding matrix at transmitter $j$ and the zero-forcing matrix at receiver $k$, respectively, while ${{\mathbf{H}}_{kj}}$ denotes the channel matrix from transmitter $j$ to receiver $k$, $j,k\in \left[ 1,K \right]$. Then, the alignment conditions can be summarized as
\begin{IEEEeqnarray}{lcr}
\mathbf{U}_{k}^{H}{{\mathbf{H}}_{kj}}{{\mathbf{V}}_{j}}=\mathbf{0},\text{ }\forall k\ne j \\ 
\operatorname{rank}\left( \mathbf{U}_{k}^{H}{{\mathbf{H}}_{kk}}{{\mathbf{V}}_{k}} \right)={{d}_{k}}     
\end{IEEEeqnarray}
where ${{d}_{k}}$ denotes the target DoF to achieve at receiver $k$. The conditions in (1) and (2) assure that the interference subspace $\sum\nolimits_{j\ne k}{{{\mathbf{H}}_{kj}}{{\mathbf{V}}_{j}}}$ is complementary to the signal subspace ${{\mathbf{H}}_{kk}}{{\mathbf{V}}_{k}}$ at the receiver $k$. A few heuristic algorithms have been proposed to solve the transceivers, $\left\{ {{\mathbf{V}}_{j}} \right\}$ and $\left\{ {{\mathbf{U}}_{k}} \right\}$, in an iterative manner [2,3]. However, these algorithms cannot determine whether the transceivers exist for a given tuple of DoFs $\left( {{d}_{1}},{{d}_{2}},\cdots,{{d}_{K}} \right)$, nor is there any guarantee for converging to the optimal transceivers even when they exist. Later, the authors in [4] have shown that it is NP-hard to find $\left\{ {{\mathbf{V}}_{j}} \right\}$ and $\left\{ {{\mathbf{U}}_{k}} \right\}$ jointly by solving (1) and (2) with a given tuple of DoFs [4]. They have also proposed an iterative algorithm to solve $\left\{ {{\mathbf{V}}_{j}} \right\}$ and $\left\{ {{\mathbf{U}}_{k}} \right\}$, which can computes a local optimal solution without resorting to a priori specification of the DoF tuple. In [4], however, with the utility function of ${SINR}/{\left( 1+SINR \right)}$, the optimized performance is not guaranteed if the interferences fail to be aligned, since it is still unknown whether the IA transceivers exist or not.

In summary, the aforementioned works have raised various issues to find the IA transceivers. Obviously, IAC inherits the NP-hardness from IA in terms of finding the transceivers for a given tuple of DoFs. In our earlier work, therefore, instead of investigating the NP-hard problem for IAC, we consider suboptimal heuristics design. By proposing a symbol-to-symbol (STS) alignment structure and constructing an IAC graph, we have derived the necessary and sufficient conditions for the existence of IAC transceivers [5]. In this paper, we aim to design the closed-form transceivers when they exist for IAC in MIMO interference channel. We first propose a general design for any given tuple of DoFs $\left( {{d}_{1}},{{d}_{2}},\cdots,{{d}_{K}} \right)$ that satisfies the proposed necessary and sufficient conditions. Then, we present how the optimal IAC transceivers can be designed to achieve the maximum DoFs of $2M$.

The remainder of this paper is organized as follows: Section II presents the system model and reviews how IAC works in MIMO interference channel. Section III presents the proposed IAC with symbol-to-symbol alignment scheme and list the main results obtained in [5]. Section IV provides the general design principle for the closed-form IAC transceivers while providing the optimum design on them in Section V. Finally, conclusion is made and future works are suggested in the last section.

\section{System Model}
Fig. 1 illustrates a MIMO interference channel, in which $K$ transceiver pairs share the same resource while each transmitter and receiver are equipped with $M$ antennas. Assume that transmitter $j$ sends ${{d}_{j}}$ independent data streams to receiver $j$, incurring interferences to other receivers $\left( {{d}_{j}}\le M,\text{ }j=1,2,\cdots ,K \right)$. We can further define the total DoF in the system as ${{\operatorname{DoF}}^{sys}}=\sum\nolimits_{j=1}^{K}{{{d}_{j}}}$. Let ${{\mathbf{x}}_{j}}$ denote the transmitted signal vectors of dimension ${{d}_{j}}\times 1$ from the transmitter $j$, in which each element of the vector corresponds to one independent data symbol, denoted by ${{x}_{j\ell }}$, $\ell =1,2,\cdots ,{{d}_{j}}$. Furthermore, let ${{\mathbf{H}}_{kj}}$ denote an $M\times M$ channel matrix from transmitter $j$ to receiver $k$ with each entry drawn independently from a continuous distribution while allowing no channel extension, and ${{\mathbf{V}}_{j}}=[{{\mathbf{v}}_{j1}},{{\mathbf{v}}_{j2}},\cdots,{{\mathbf{v}}_{j{{d}_{j}}}}]$ of dimension $M\times {{d}_{j}}$ represent a transmit precoding matrix at transmitter $j$ where each column vector is applied to each data symbol $\left( j,k=1,2,\cdots,K \right)$. Meanwhile, let ${{\mathbf{U}}_{k}}$ of dimension $M\times {{d}_{k}}$ be a zero-forcing matrix at receiver $k$ and ${{\mathbf{y}}_{k}}$ denote the output signal vector of dimension ${{d}_{k}}\times 1$ at receiver $k$, which is given by
\begin{IEEEeqnarray}{lcr}
\begin{array}{l}
{{\mathbf{y}}_{k}}=\underbrace{\mathbf{U}_{k}^{H}{{\mathbf{H}}_{kk}}{{\mathbf{V}}_{k}}{{\mathbf{x}}_{k}}}_{\text{desired signal}}+\underbrace{\mathbf{U}_{k}^{H}\sum\limits_{j=1;j\ne k}^{K}{{{\mathbf{H}}_{kj}}{{\mathbf{V}}_{j}}{{\mathbf{x}}_{j}}}}_{\text{interference}}+\mathbf{U}_{k}^{H}{{\mathbf{n}}_{k}},$ $$ $$ $
\\ $ $$ $$ $$ $$ $$ $$ $$ $$ $$ $$ $$ $$ $$ $$ $$ $$ $$ $$ $$ $$ $$ $$ $$ $$ $$ $$ $$ $$ $$ $$ $$ $$ $$ $$ $$ $$ $$ $$ $$ $$ $$ $$ $$ $$ $$ $k=1,2,\cdots ,K
\end{array}
\end{IEEEeqnarray}
where ${{\mathbf{n}}_{k}} \sim N\left( \mathbf{0},{{\sigma }^{2}}\mathbf{I} \right)$ is zero-mean additive white Gaussian noise and ${{\mathbf{A}}^{H}}$ denotes a conjugate transpose of a matrix $\mathbf{A}$.

As each data symbol ${{x}_{j\ell }}$ is encoded by using a precoding vector ${{\mathbf{v}}_{j\ell }}$ and received through the channel as ${{\mathbf{H}}_{kj}}{{\mathbf{v}}_{j\ell }}{{x}_{j\ell }}$ at receiver $k$, the signal subspace ${{\tilde{S}}_{k}}$ can be expressed by a set of signal vectors, ${{\mathcal{S}}_{k}}=\left\{ {{\mathbf{H}}_{kk}}{{\mathbf{v}}_{k\ell }} \right\}$ for $\ell \in \left[ 1,{{d}_{k}} \right]$, while the interference subspace ${{\tilde{I}}_{k}}$ can be expressed by a set of interference vectors ${{\mathcal{I}}_{k}}=\left\{ {{\mathbf{H}}_{kj}}{{\mathbf{v}}_{j\ell }} \right\}$ for $j\in \left[ 1,K \right]$, $j\ne k$, and $\ell \in \left[ 1,{{d}_{j}} \right]$. Considering that IAC is a combined scheme of IA and IC, ${{\mathcal{I}}_{k}}$ can be also written into a union of two sets, ${{\mathcal{I}}_{k}}=\mathcal{I}_{k}^{\operatorname{IA}}\bigcup \mathcal{I}_{k}^{\operatorname{IC}}$, where $\mathcal{I}_{k}^{\operatorname{IA}}$ and $\mathcal{I}_{k}^{\operatorname{IC}}$ denote the sets of interference vectors to be mitigated by IA and IC operations, respectively.

At the receive side, IC operation works to subtract interference effect caused by the known signals, i.e., the decoded signals ${{\mathbf{\hat{x}}}_{k}}$ at receiver $k$ are sent over a backhaul link to other receivers that have not performed the decoding operation yet, so that the interference effect caused by ${{\mathbf{x}}_{k}}$ can be cancelled. For the purpose of successive cancellation, therefore, a decoding operation is performed in one receiver at a time. Without loss of generality, we assume that the decoding order follows from receiver $1$ to $K$ in the subsequent discussion. As receiver $k$ can obtain the decoded signals $\left\{ {{{\mathbf{\hat{x}}}}_{j}} \right\}_{j=1}^{k-1}$ from receivers $1$ to $\left( k-1 \right)$ through the backhaul link, an estimate of the corresponding interference $\sum\nolimits_{j=1}^{k-1}{{{\mathbf{H}}_{kj}}{{\mathbf{V}}_{j}}{{\mathbf{x}}_{j}}}$ can be cancelled from ${{\mathbf{y}}_{k}}$, $k=2,3,\cdots ,K$. As a result, a set of interference vectors mitigated by IC operation can be expressed by $\mathcal{I}_{k}^{\operatorname{IC}}=\left\{ {{\mathbf{H}}_{kj}}{{\mathbf{v}}_{j\ell }} \right\}$ for $j\in \left[ 1,k-1 \right]$ and $\ell \in \left[ 1,{{d}_{j}} \right]$.

\begin{figure}[!t]
\centering
\includegraphics[width=3.4in]{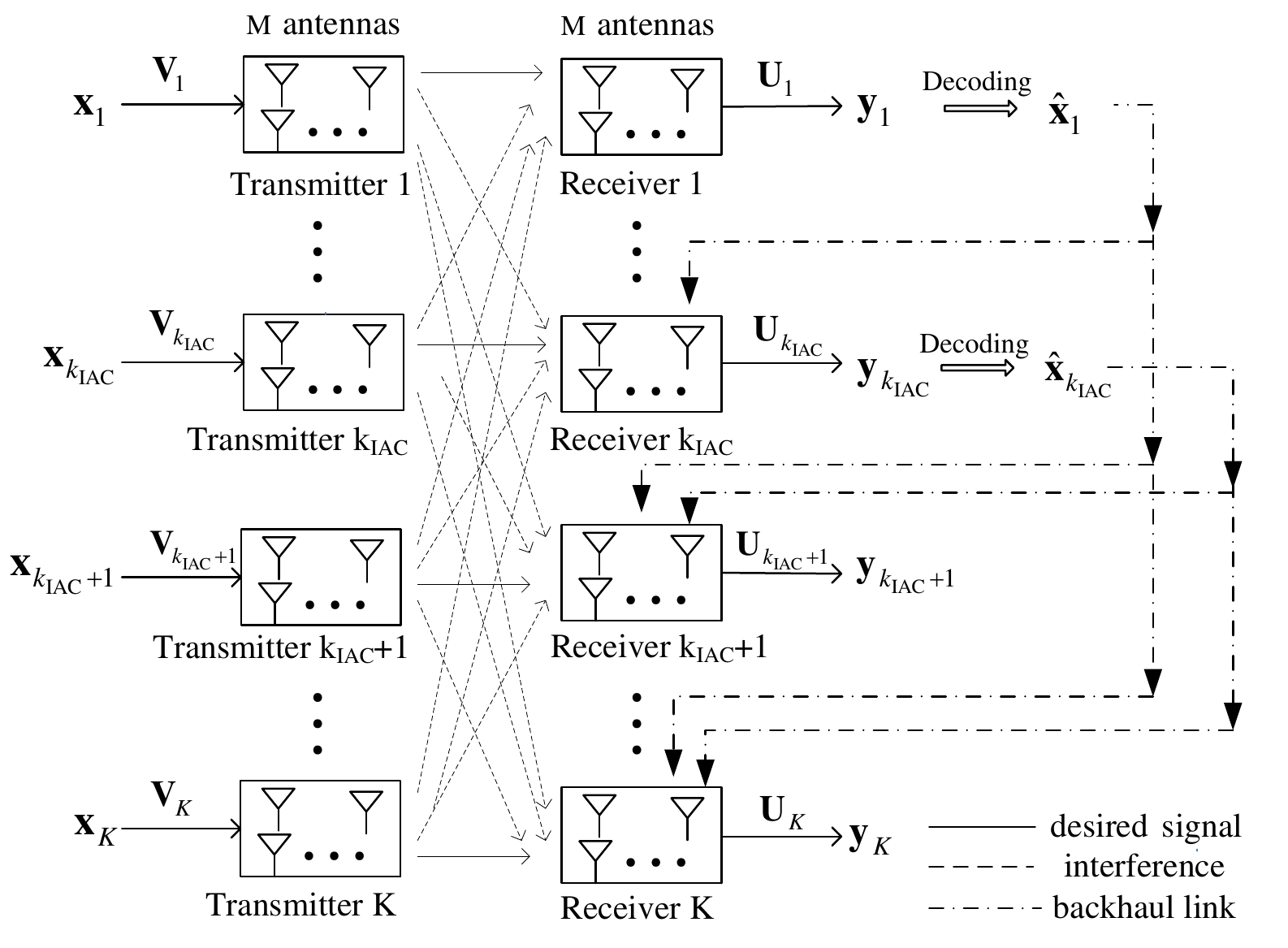}
\vspace*{-10pt}
\caption{$K\times K$ MIMO interference channel model for IAC}
\end{figure}

At the transmit side, IA operation works to precode the signals jointly, so that the interferences incurred at the receive side can be effectively aligned. As IC operation can directly subtract the interference signals in $\mathcal{I}_{k}^{\operatorname{IC}}$ at receiver $k$, such a part of interference is not considered for IA operation. In other words, IA operation is responsible for only the other part of interference which is not cancelled, i.e., we have $\mathcal{I}_{k}^{\operatorname{IA}}=\left\{ {{\mathbf{H}}_{kj}}{{\mathbf{v}}_{j\ell }} \right\}$ for $j\in \left[ k+1,K \right]$ and $\ell \in \left[ 1,{{d}_{j}} \right]$. Additionally, after a process of successive cancellation, the number of the remaining received symbols at each receiver $k$, given as $\left| {{\mathcal{S}}_{k}} \right|+\left| \mathcal{I}_{k}^{\operatorname{IA}} \right|=\sum\nolimits_{j=k}^{K}{{{d}_{j}}}$, decreases as the cancellation process progresses in order. As long as $\sum\nolimits_{j=k}^{K}{{{d}_{j}}}\le M$, the desired signals can be decoded without resorting to IA operation because a total of no more than $M$ packets can be separated by $M$ antennas. With $\left| {{\mathcal{S}}_{k}} \right|+\left| \mathcal{I}_{k}^{\operatorname{IA}} \right|\le M$ for receiver $k$, it is always true that $\left| {{\mathcal{S}}_{j}} \right|+\left| \mathcal{I}_{j}^{\operatorname{IA}} \right|\le M$ for $j\in \left[ k+1,K \right]$ and therefore, all remaining receivers $\left( k+1 \right),\left( k+2 \right),\cdots,K$ can be also decoded directly without resorting to IA operation. Therefore, the decoded signals $\left\{ {{{\mathbf{\hat{x}}}}_{j}} \right\}_{j=k}^{K}$ are not required to be shared for a successive cancellation purpose. If receiver $k$ is the first receiver which satisfies $\left| {{\mathcal{S}}_{k}} \right|+\left| \mathcal{I}_{k}^{\operatorname{IA}} \right|\le M$, a new label ${{k}_{\operatorname{IAC}}}$ is allocated to receiver $\left( k-1 \right)$ so as to indicate that only the interference received at receiver $1$ to ${{k}_{\operatorname{IAC}}}$ is required to be aligned and furthermore, only the decoded signals $\left\{ {{{\mathbf{\hat{x}}}}_{j}} \right\}_{j=1}^{{{k}_{\operatorname{IAC}}}}$ have to be shared through the backhaul link for a cancellation purpose.

Consequently, two steps are required for decoding in IAC scheme. In the first step, receivers $1$ to ${{k}_{\operatorname{IAC}}}$ will perform the decoding operation with one receiver at a time and then, the decoded signals $\left\{ {{{\mathbf{\hat{x}}}}_{j}} \right\}_{j=1}^{{{k}_{\operatorname{IAC}}}}$ are shared with receivers $\left( j+1 \right)$ to $K$ through a backhaul link for cancellation. In the second step, all other receivers, indexed by $({{k}_{\operatorname{IAC}}}+1),({{k}_{\operatorname{IAC}}}+2),\cdots,$ and $K$, decode their own signals simultaneously. Note that additional process of sending the decoded packets over the backhaul link in the first step is considered as overhead. For simplicity, here we roughly measure the corresponding overhead by counting the number of receivers, to which a decoded packet has to be sent over the backhaul link. For example, if ${{x}_{1\ell }}$ has been decoded at receiver 1, it would be sent to all other  $\left( K-1 \right)$ receivers. Thus, the overhead associated with ${{x}_{1\ell }}$ is counted as $\left( K-1 \right)$ packets. Similarly, ${{x}_{j\ell }}$ will have the overhead of $\left( K-j \right)$ packets. Note that its overhead is zero for $j>{{k}_{\text{IAC}}}$. Let ${{O}_{\operatorname{sys}}}$ denote the total overhead associated with the whole system, given as ${{O}_{\operatorname{sys}}}=\sum\nolimits_{j=1}^{{{k}_{\operatorname{IAC}}}}{\left( K-j \right){{d}_{j}}}$.

\section{Interference Alignment and Cancellation with Symbol-to-Symbol Alignment Scheme}
In our earlier work [5], instead of investigating the NP-hard problem for a joint transceiver design, we consider a design of the transmitter $\left\{ {{\mathbf{V}}_{j}} \right\}$ first and then the receiver $\left\{ {{\mathbf{U}}_{k}} \right\}$ for IAC. Aiming for a successful decoding at each receiver $k$, the transmitter $\left\{ {{\mathbf{V}}_{j}} \right\}$ should be designed so that the signal subspace ${{\tilde{S}}_{k}}$ can be complimentary to the interference subspace ${{\tilde{I}}_{k}}$. This indicates two constraints to meet: first, ${{\tilde{S}}_{k}}$ and ${{\tilde{I}}_{k}}$ should be linearly independent; second, the dimension of ${{\tilde{S}}_{k}}$ and ${{\tilde{I}}_{k}}$ should satisfy the following condition:
\begin{IEEEeqnarray}{lcr}
\dim({{\tilde{S}}_{k}})+\dim({{\tilde{I}}_{k}})\le M              
\end{IEEEeqnarray}             
where $\dim\left( \mathcal{A} \right)$ corresponds to the cardinality of a basis for a vector space $\mathcal{A}$. As channel matrices $\{{{\mathbf{H}}_{kj}}\}$ are given, ${{\tilde{S}}_{k}}$ and ${{\tilde{I}}_{k}}$ can be determined as long as the transmitter $\left\{ {{\mathbf{V}}_{j}} \right\}$ are solved, and then, the receiver $\left\{ {{\mathbf{U}}_{k}} \right\}$ can always be found in the left null space of ${{\tilde{I}}_{k}}$. In order to meet the two constraints given in the above for $\left\{ {{\mathbf{V}}_{j}} \right\}$, we have proposed a symbol-to-symbol (STS) alignment structure for IAC in [5], based on which an IAC graph has been constructed so that the necessary and sufficient conditions for the existence of closed-form transceiver solutions can be derived.

In this section, we first briefly review the STS alignment structure. Then, we give an illustrative example to show how IAC with STS alignment scheme works in MIMO interference channel. At last, a notion of IAC graph is reviewed and then, the associated results are also discussed.

\subsection{Symbol-to-symbol (STS) alignment structure: Overview}
The constraint on ${{\mathbf{V}}_{j}}$ in (4) motivates us to align ${{\tilde{I}}_{k}}$ onto a set of basis vectors with dimension ${{Z}_{k}}=M\text{-}\dim({{{\tilde{S}}}_{k}})$. As the interference vectors in $\mathcal{I}_{k}^{\operatorname{IC}}$ can be cancelled directly and do not burden on the dimension of ${{\tilde{I}}_{k}}$, only the interference vectors in $\mathcal{I}_{k}^{\operatorname{IA}}$ are considered for alignment. The STS alignment structure can be constructed by two steps: first, a set of basis vectors, denoted by $\mathcal{\bar{I}}_{k}^{\operatorname{SIA}}$, is directly {\it selected} among interference vectors in $\mathcal{I}_{k}^{\operatorname{IA}}$, i.e., $\mathcal{\bar{I}}_{k}^{\operatorname{SIA}}=\left\{ {{{\mathbf{\bar{i}}}}_{1,k}},\cdots,{{{\mathbf{\bar{i}}}}_{n,k}},\cdots,{{{\mathbf{\bar{i}}}}_{{{Z}_{k}},k}} \right\}\subset \mathcal{I}_{k}^{\operatorname{IA}}$ , where the superscription ‘SIA’ indicates the STS interference alignment, while the remaining interference vectors $\{{{\mathbf{i}}_{f,k}}\}_{f=1}^{_{\left| {{\mathcal{I}}_{k}} \right|-{{Z}_{k}}}}$ belong to $\left( \mathcal{I}_{k}^{\operatorname{IA}}-\mathcal{\bar{I}}_{k}^{\operatorname{SIA}} \right)$, i.e., $\mathcal{I}_{k}^{\operatorname{IA}}-\mathcal{\bar{I}}_{k}^{\operatorname{SIA}}=\left\{ {{\mathbf{i}}_{1,k}},\cdots,{{\mathbf{i}}_{f,k}},\cdots,{{\mathbf{i}}_{\left| \mathcal{I}_{k}^{\operatorname{IA}} \right|-{{Z}_{k}},k}} \right\}\subset \mathcal{I}_{k}^{\operatorname{IA}}$; second, alignment is implemented between single vectors corresponding to one symbol each. Once $\mathcal{\bar{I}}_{k}^{\operatorname{SIA}}$ is determined in the first step, each remaining interference vector, ${{\mathbf{i}}_{f,k}}\in \mathcal{I}_{k}^{\operatorname{IA}}-\mathcal{\bar{I}}_{k}^{\operatorname{SIA}}$, will be aligned onto only one of the basis vectors, ${{\mathbf{\bar{i}}}_{n,k}}\in \mathcal{\bar{I}}_{k}^{\operatorname{SIA}}$, i.e., $\operatorname{span}\left( {{\mathbf{i}}_{f,k}} \right)=\operatorname{span}\left( {{{\mathbf{\bar{i}}}}_{n,k}} \right)$. Each ${{\mathbf{i}}_{f,k}}$ can be pick out only once while ${{\mathbf{\bar{i}}}_{n,k}}$ can be employed repeatedly. Replacing ${{\mathbf{i}}_{f,k}}$ and ${{\mathbf{\bar{i}}}_{n,k}}$ with the detailed expression, alignment equations yielded from the STS alignment structure for user $k$ can be represented as
\begin{IEEEeqnarray}{lcr}
\begin{array}{l}
{{\Phi }_{k}} \triangleq \{ \left. \operatorname{span}( {{\mathbf{H}}_{kj}}{{\mathbf{v}}_{j\ell }} )=\operatorname{span}( {{\mathbf{H}}_{kj'}}{{\mathbf{v}}_{j'{\ell }'}} ) \right| $ $$ $$ $$ $$ $$ $$ $$ $$ $$ $$ $$ $$ $$ $$ $$ $$ $$ $$ $
\\ $ $$ $$ $$ $$ ${{\mathbf{H}}_{kj'}}{{\mathbf{v}}_{j'{\ell }'}}\in \mathcal{I}_{k}^{\operatorname{IA}}-\mathcal{\bar{I}}_{k}^{\operatorname{SIA}},\text{ }{{\mathbf{H}}_{kj}}{{\mathbf{v}}_{j\ell }}\in \mathcal{\bar{I}}_{k}^{\operatorname{SIA}},\text{ }j\ne {j}' \},
\\ $ $$ $$ $$ $$ $$ $$ $$ $$ $$ $$ $$ $$ $$ $$ $$ $$ $$ $$ $$ $$ $$ $$ $$ $$ $$ $$ $$ $$ $$ $$ $$ $$ $$ $$ $$ $$ $$ $$ $$ $$ $$ $$ $k=1,2,\cdots ,{{k}_{\operatorname{IAC}}}
\end{array}
\end{IEEEeqnarray} 
where $j\ne {j}'$ indicates that the interferences coming from the same user cannot be aligned [7]. Then, a complete set of linear alignment equations yielded from the STS alignment structure for a whole system is given by $\Psi \triangleq \left\{ \left. {{\Phi }_{k}} \right|k\in \left[ 1,{{k}_{\operatorname{IAC}}} \right] \right\}$.

\begin{figure}[!t]
\centering
\includegraphics[width=3.4in]{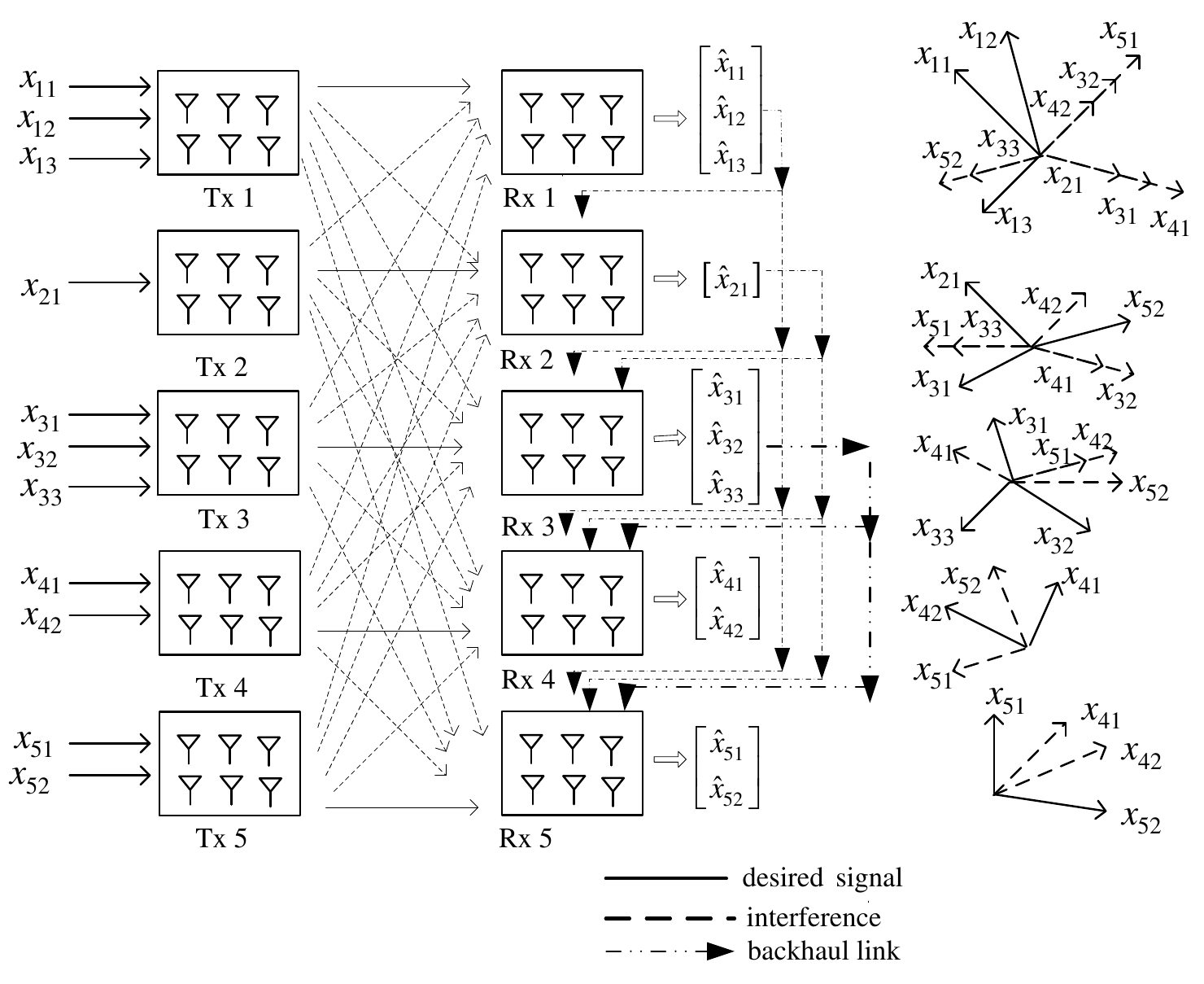}
\vspace*{-10pt}
\caption{IAC example with $M=6$, ${{d}_{1}}={{d}_{3}}=3$, ${{d}_{2}}=1$ and ${{d}_{4}}={{d}_{5}}=2$ for $K=5$: ${{k}_{\operatorname{IAC}}}=3$}
\end{figure}

\subsection{Illustrative example}
To illustrate how IAC with the STS alignment works, let us consider an example for $K=5$ and $M=6$ in Fig. 2, where we assume ${{d}_{1}}={{d}_{3}}=3$, ${{d}_{2}}=1$, and ${{d}_{4}}={{d}_{5}}=2$, i.e., transmitting a total number of eleven data symbols at the same time $\left( \sum\nolimits_{j=1}^{5}{{{d}_{j}}}=11 \right)$. As $\sum\nolimits_{j=3}^{5}{{{d}_{j}}}=7>M$ and $\sum\nolimits_{j=4}^{5}{{{d}_{j}}}=4<M$, ${{k}_{\operatorname{IAC}}}=3$. It implies that the receivers 4 and 5 can decode ${{\mathbf{x}}_{4}}$ and ${{\mathbf{x}}_{5}}$ without cancellation. In the sequel, we present a set of alignment equations $\Psi =\left\{ {{\Phi }_{j}}|j=1,2,\cdots ,{{k}_{\operatorname{IAC}}} \right\}$ for this example. At receiver 1, three desired data symbols ${{x}_{11}}$, ${{x}_{12}}$, and ${{x}_{13}}$ will be decoded and thus, $\dim({{{\tilde{S}}}_{1}})=3$. In order to meet (4), all other eight symbols, $\left\{ {{x}_{2{{\ell }_{2}}}} \right\}_{{{\ell }_{2}}=1}^{{{d}_{2}}}$ , $\left\{ {{x}_{3{{\ell }_{3}}}} \right\}_{{{\ell }_{3}}=1}^{{{d}_{3}}}$, $\left\{ {{x}_{4{{\ell }_{4}}}} \right\}_{{{\ell }_{4}}=1}^{{{d}_{4}}}$, and $\left\{ {{x}_{5{{\ell }_{5}}}} \right\}_{{{\ell }_{5}}=1}^{{{d}_{5}}}$, considered as interferences, have to be aligned so that $\dim({{{\tilde{I}}}_{1}})\le M\text{-}\dim({{{\tilde{S}}}_{1}})=3$. To construct a set of basis vectors of ${{\tilde{I}}_{1}}$, let us select the interference vectors, ${{\mathbf{H}}_{12}}{{\mathbf{v}}_{21}}$, ${{\mathbf{H}}_{13}}{{\mathbf{v}}_{33}}$, and ${{\mathbf{H}}_{14}}{{\mathbf{v}}_{42}}$, out of  $\mathcal{I}_{1}^{\operatorname{IA}}$, which is represented by $\mathcal{\bar{I}}_{1}^{\operatorname{SIA}}=\left\{ {{\mathbf{H}}_{12}}{{\mathbf{v}}_{21}},{{\mathbf{H}}_{13}}{{\mathbf{v}}_{33}},{{\mathbf{H}}_{14}}{{\mathbf{v}}_{42}} \right\}$. Then we have $\mathcal{I}_{1}^{\operatorname{IA}}-\mathcal{\bar{I}}_{1}^{\operatorname{SIA}}=\left\{ {{\mathbf{H}}_{13}}{{\mathbf{v}}_{31}},{{\mathbf{H}}_{13}}{{\mathbf{v}}_{32}},{{\mathbf{H}}_{14}}{{\mathbf{v}}_{41}},{{\mathbf{H}}_{15}}{{\mathbf{v}}_{51}},{{\mathbf{H}}_{15}}{{\mathbf{v}}_{52}} \right\}$. Following the STS structure, each interference vector in $\left( \mathcal{I}_{1}^{\operatorname{IA}}-\mathcal{\bar{I}}_{1}^{\operatorname{SIA}} \right)$ will be uniquely aligned onto one of the basis vectors in $\mathcal{\bar{I}}_{1}^{\operatorname{SIA}}$. For example, assume that ${{\mathbf{H}}_{13}}{{\mathbf{v}}_{31}}$ and ${{\mathbf{H}}_{14}}{{\mathbf{v}}_{41}}$ are aligned onto ${{\mathbf{H}}_{12}}{{\mathbf{v}}_{21}}$; ${{\mathbf{H}}_{15}}{{\mathbf{v}}_{52}}$ is aligned onto ${{\mathbf{H}}_{13}}{{\mathbf{v}}_{33}}$; and ${{\mathbf{H}}_{13}}{{\mathbf{v}}_{32}}$ and ${{\mathbf{H}}_{15}}{{\mathbf{v}}_{51}}$ are aligned onto ${{\mathbf{H}}_{14}}{{\mathbf{v}}_{42}}$, constructing the following set of alignment equations, ${{\Phi }_{1}}$, for the receiver 1:
\begin{IEEEeqnarray}{lcr}
\begin{array}{l}
\operatorname{span}({{\mathbf{H}}_{12}}{{\mathbf{v}}_{21}})=\operatorname{span}({{\mathbf{H}}_{13}}{{\mathbf{v}}_{31}} ) \\
\operatorname{span}({{\mathbf{H}}_{12}}{{\mathbf{v}}_{21}})=\operatorname{span}({{\mathbf{H}}_{14}}{{\mathbf{v}}_{41}}) \\
\operatorname{span}({{\mathbf{H}}_{1\text{3}}}{{\mathbf{v}}_{\text{33}}})=\operatorname{span}({{\mathbf{H}}_{15}}{{\mathbf{v}}_{52}}) \\
\operatorname{span}({{\mathbf{H}}_{14}}{{\mathbf{v}}_{42}})=\operatorname{span}({{\mathbf{H}}_{13}}{{\mathbf{v}}_{32}}) \\
\operatorname{span}({{\mathbf{H}}_{14}}{{\mathbf{v}}_{42}})=\operatorname{span}({{\mathbf{H}}_{15}}{{\mathbf{v}}_{51}})
\end{array}
\end{IEEEeqnarray} 
After decoding ${{\hat{x}}_{11}}$, ${{\hat{x}}_{12}}$, and ${{\hat{x}}_{13}}$, they would be sent to the receivers 2, 3, 4, and 5 for a successive cancellation purpose.

At receiver 2, since ${{\hat{x}}_{11}}$, ${{\hat{x}}_{12}}$, and ${{\hat{x}}_{13}}$ from receiver 1 can be subtracted by IC operation, we have $\mathcal{I}_{2}^{\operatorname{IC}}=\left\{ \left. {{\mathbf{H}}_{21}}{{\mathbf{v}}_{1{{\ell }_{1}}}} \right|{{\ell }_{1}}\in \left[ 1,{{d}_{1}} \right] \right\}$ and $\mathcal{I}_{2}^{\operatorname{IA}}=\left\{ \left. {{\mathbf{H}}_{2j}}{{\mathbf{v}}_{j{{\ell }_{j}}}} \right|j\in \left[ 3,5 \right],\text{ }{{\ell }_{j}}\in \left[ 1,{{d}_{j}} \right] \right\}$. With $\dim({{\tilde{I}}_{2}})\le M\text{-}\dim({{\tilde{S}}_{2}})=5$, we select five interference vectors in $\mathcal{I}_{2}^{\operatorname{IA}}$ to construct $\mathcal{\bar{I}}_{2}^{\operatorname{SIA}}$, given as $\mathcal{\bar{I}}_{2}^{\operatorname{SIA}}=\left\{ {{\mathbf{H}}_{23}}{{\mathbf{v}}_{31}},{{\mathbf{H}}_{23}}{{\mathbf{v}}_{33}},{{\mathbf{H}}_{24}}{{\mathbf{v}}_{41}},{{\mathbf{H}}_{24}}{{\mathbf{v}}_{42}},{{\mathbf{H}}_{25}}{{\mathbf{v}}_{52}} \right\}$. Following the STS structure again, for example, ${{\Phi }_{2}}$ can be constructed with the following alignment equations:
\begin{IEEEeqnarray}{lcr}
\begin{array}{l}
\operatorname{span}({{\mathbf{H}}_{24}}{{\mathbf{v}}_{41}})=\operatorname{span}({{\mathbf{H}}_{23}}{{\mathbf{v}}_{32}}) \\
\operatorname{span}({{\mathbf{H}}_{23}}{{\mathbf{v}}_{33}})=\operatorname{span}({{\mathbf{H}}_{25}}{{\mathbf{v}}_{51}})
\end{array}
\end{IEEEeqnarray} 
As ${{\hat{x}}_{21}}$, ${{\hat{x}}_{22}}$, and ${{\hat{x}}_{23}}$ are decoded now in the receiver 2, they will be shared with the receivers 3, 4, and 5 for cancellation.

At receiver 3, we have $\mathcal{I}_{3}^{\operatorname{IC}}=\left\{ \left. {{\mathbf{H}}_{3j}}{{\mathbf{v}}_{j{{\ell }_{j}}}} \right|j\in \left[ 1,2 \right],\text{ }{{\ell }_{j}}\in \left[ 1,{{d}_{j}} \right] \right\}$ and $\mathcal{I}_{3}^{\operatorname{IA}}=\left\{ \left. {{\mathbf{H}}_{3j}}{{\mathbf{v}}_{j{{\ell }_{j}}}} \right|j\in \left[ 4,5 \right],\text{ }{{\ell }_{j}}\in \left[ 1,{{d}_{j}} \right] \right\}$. With $\dim({{\tilde{I}}_{3}})\le M\text{-}\dim({{\tilde{S}}_{3}})=3$, we select three interference vectors in $\mathcal{I}_{3}^{\operatorname{IA}}$ to construct $\mathcal{\bar{I}}_{3}^{\operatorname{SIA}}$, e.g., $\mathcal{\bar{I}}_{3}^{\operatorname{SIA}}=\left\{ {{\mathbf{H}}_{34}}{{\mathbf{v}}_{41}},{{\mathbf{H}}_{35}}{{\mathbf{v}}_{51}},{{\mathbf{H}}_{35}}{{\mathbf{v}}_{52}} \right\}$. As $\left( \mathcal{I}_{3}^{\operatorname{IA}}-\mathcal{\bar{I}}_{3}^{\operatorname{SIA}} \right)$ has only one interference vector, it ends up with only one alignment equation, i.e., $\left| {{\Phi }_{3}} \right|=\left| \mathcal{I}_{3}^{IA}-\mathcal{\bar{I}}_{3}^{\operatorname{SIA}} \right|=1$, where ${{\Phi }_{3}}=\left\{ \operatorname{span}({{\mathbf{H}}_{35}}{{\mathbf{v}}_{51}})=\operatorname{span}({{\mathbf{H}}_{34}}{{\mathbf{v}}_{42}}) \right\}$.

As ${{k}_{\operatorname{IAC}}}$ = 3 in our example, the receivers 4 and 5 can decode their desired symbols without resorting to alignment operation. Consequently, we have obtained a complete set of linear alignment equations for user $k\in [1,{{k}_{\operatorname{IAC}}}]$, $\Psi =\left\{ {{\Phi }_{1}},{{\Phi }_{2}},{{\Phi }_{3}} \right\}$. With some algebraic manipulations, $\Psi$ can be equivalently written as
\begin{IEEEeqnarray}{lcr}
\begin{array}{l}
{{\mathbf{v}}_{21}}={{\left( {{\mathbf{H}}_{12}} \right)}^{-1}}{{\mathbf{F}}_{1}}{{\mathbf{v}}_{42}} \\
{{\mathbf{v}}_{31}}={{\left( {{\mathbf{H}}_{13}} \right)}^{-1}}{{\mathbf{F}}_{1}}{{\mathbf{v}}_{42}} \\
{{\mathbf{v}}_{32}}={{\left( {{\mathbf{H}}_{13}} \right)}^{-1}}{{\mathbf{H}}_{14}}{{\mathbf{v}}_{42}} \\
{{\mathbf{v}}_{33}}={{\mathbf{F}}_{\text{2}}}{{\mathbf{v}}_{42}} \\
{{\mathbf{v}}_{41}}={{\left( {{\mathbf{H}}_{24}} \right)}^{-1}}{{\mathbf{H}}_{23}}{{\left( {{\mathbf{H}}_{13}} \right)}^{-1}}{{\mathbf{H}}_{14}}{{\mathbf{v}}_{42}} \\
{{\mathbf{v}}_{51}}={{\left( {{\mathbf{H}}_{35}} \right)}^{-1}}{{\mathbf{H}}_{34}}{{\mathbf{v}}_{42}} \\
{{\mathbf{v}}_{52}}={{\left( {{\mathbf{H}}_{15}} \right)}^{-1}}{{\mathbf{H}}_{1\text{3}}}{{\mathbf{F}}_{\text{2}}}{{\mathbf{v}}_{42}} \\
\operatorname{span}({{\mathbf{v}}_{42}})=\operatorname{span}({{\mathbf{F}}_{\text{3}}}{{\mathbf{v}}_{42}})
\end{array}
\end{IEEEeqnarray} 
where ${{\mathbf{F}}_{1}}={{\mathbf{H}}_{14}}{{\left( {{\mathbf{H}}_{24}} \right)}^{-1}}{{\mathbf{H}}_{23}}{{\left( {{\mathbf{H}}_{13}} \right)}^{-1}}{{\mathbf{H}}_{14}}$, ${{\mathbf{F}}_{2}}={{\left( {{\mathbf{H}}_{23}} \right)}^{-1}}{{\mathbf{H}}_{25}}{{\left( {{\mathbf{H}}_{35}} \right)}^{-1}}{{\mathbf{H}}_{34}}$, and ${{\mathbf{F}}_{3}}={{\left( {{\mathbf{H}}_{34}} \right)}^{-1}}{{\mathbf{H}}_{35}}{{\left( {{\mathbf{H}}_{15}} \right)}^{-1}}{{\mathbf{H}}_{14}}$. Since $\operatorname{span}\left( {{\mathbf{v}}_{42}} \right)=\operatorname{span}\left( {{\mathbf{F}}_{3}}{{\mathbf{v}}_{42}} \right)$, we can set ${{\mathbf{v}}_{42}}$ to be one of eigenvectors of ${{\mathbf{F}}_{3}}$. Then, the other precoding vectors $\left\{ {{\mathbf{v}}_{2{{\ell }_{2}}}} \right\}_{{{\ell }_{2}}=1}^{{{d}_{2}}}$, $\left\{ {{\mathbf{v}}_{3{{\ell }_{3}}}} \right\}_{{{\ell }_{3}}=1}^{{{d}_{3}}}$, ${{\mathbf{v}}_{41}}$, and $\left\{ {{\mathbf{v}}_{5{{\ell }_{5}}}} \right\}_{{{\ell }_{5}}=1}^{{{d}_{5}}}$ can be found subsequently by (8).

\begin{figure}[!t]
\centering
\includegraphics[width=2.6in]{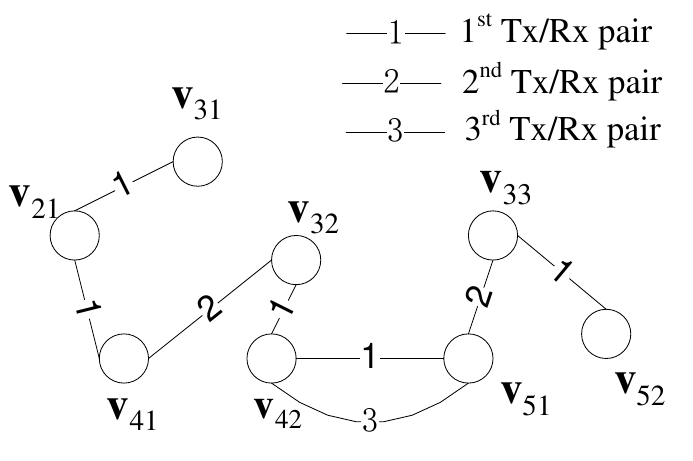}
\vspace*{-10pt}
\caption{The IAC graph corresponding to the example in Fig. 2}
\end{figure}

\subsection{IAC graph}
In order to derive the conditions for solving ${{\mathbf{V}}_{j}}$ out of $\Psi$ for a general tuple of DoFs $\left( {{d}_{1}},{{d}_{2}},\cdots,{{d}_{K}} \right)$, we have constructed an IAC graph $\mathcal{G}=\left( \mathcal{P},\mathcal{E} \right)$ in our earlier work [5], where $\mathcal{P}$ and $\mathcal{E}$ represent a set of vertices and a set of edges, respectively. We have proven that $\mathcal{G}$ and $\Psi$ have an one-to-one correspondence, i.e., each precoding vector ${{\mathbf{v}}_{j{{\ell }_{j}}}}$ and each alignment equation in $\Psi$ can be represented by one unique vertex and one unique edge in $\mathcal{G}$, respectively, and vice versa. Each edge in the IAC graph holds a label to declare the index of Tx/Rx pair over which the alignment operation is performed.

As each symbol corresponding to ${{\mathbf{i}}_{f,k}}\in \mathcal{I}_{k}^{\operatorname{IA}}-\mathcal{\bar{I}}_{k}^{\operatorname{SIA}}$ is aligned onto one of unique symbols corresponding to $\left( {{\mathbf{\bar{i}}}_{n,k}}\in \mathcal{I}_{k}^{\operatorname{IA}} \right)$ at receiver $k$, each variable may not appear in all alignment equations. Therefore, we can collect the equations that involves the same subset of variables into one subset. In other words, $\Psi$ can be divided into several independent subsets and thus, solving the independent subsets respectively is equivalent to solving $\Psi$. Correspondingly, each independent subset of $\Psi$ forms one independent connected subgraph of $\mathcal{G}$. Each connected subgraph has neither an isolated vertex, nor connection to other subgraphs. Assuming that there are $Q$ connected subgraphs, each of them denoted as  ${{\mathcal{G}}_{q}}=\left( {{\mathcal{P}}_{q}},{{\mathcal{E}}_{q}} \right)$, $q=1,2,\cdots,Q$, such that $\mathcal{G}=\left\{ {{\mathcal{G}}_{1}},{{\mathcal{G}}_{2}},\cdots,{{\mathcal{G}}_{Q}} \right\}$, $\left| \mathcal{P} \right|=\sum\nolimits_{q=1}^{Q}{\left| {{\mathcal{P}}_{q}} \right|}$, and $\left| \mathcal{E} \right|=\sum\nolimits_{q=1}^{Q}{\left| {{\mathcal{E}}_{q}} \right|}$, where ${{\mathcal{P}}_{q}}$ and ${{\mathcal{E}}_{q}}$ represent a set of vertices and a set of edges in subgraph $q$, respectively.

To give an intuitive illustration on the structure of IAC graph $\mathcal{G}$, let us revisit the example in Subsection III (B). With $\Psi $ given in (8), the IAC graph $\mathcal{G}$ can be depicted accordingly, as shown in Fig. 3. We that there exist only one subgraph in this example, namely, $\mathcal{G}=\left\{ {{\mathcal{G}}_{1}} \right\}$, and $\left| \mathcal{P} \right|=\left| \mathcal{E} \right|=8$. Furthermore, the label of each edge is consistent with the index of Tx/Rx pair that is subject to the alignment operation. Note that there exists only one loop in ${{\mathcal{G}}_{1}}$ in this example.

By analyzing the IAC graph $\mathcal{G}$, the necessary and sufficient conditions for solving $\Psi$ has been derived in [5], which can be summarized by the following proposition:
\begin{proposition}                                                       
{\it For a connected subgraph in IAC graph $\mathcal{G}$, if and only if the vertices form at most one loop, the precoding vectors involved can always be solved.}
\end{proposition}

Since there exists only one subgraph and only loop formed in the IAC graph of Fig. 3,  it is clear by Proposition 1 that the set of linear alignment equations $\Psi$ in (8) is solvable. Then, it has been further proven that the solutions $\{{{\mathbf{v}}_{j\ell }}\}$, $j\in \left[ 1,K \right]$ and $\ell \in \left[ 1,{{d}_{j}} \right]$, obtained from $\Psi$, can always guarantee the independence of ${{\tilde{S}}_{k}}$ and ${{\tilde{I}}_{k}}$. Finally, the necessary and sufficient conditions for the existence of closed-form transceivers for IAC with the STS alignment has been proven for $\left( K,M,J \right)$ Gaussian interference MAC system in [5], where $J$ is the total number of users. For $J=K$, $\left( K,M,J \right)$ MAC channel reduces to the $K$-user MIMO channel which is considered in this paper. The necessary and sufficient conditions accordingly can be stated by the following theorem:
\begin{theorem}
{\it In $K$-user MIMO interference channel, for a given tuple of DoF $\left( {{d}_{1}},{{d}_{2}},\cdots,{{d}_{K}} \right)$, the closed-form transceivers for IAC with STS alignment exist if and only if the following inequalities are satisfied:}
\end{theorem}
\begin{IEEEeqnarray}{lcr}
$ $$ ${{d}_{k}}+\max \{{{d}_{k+1}},{{d}_{k+2}},\cdots,{{d}_{K}}\}\le M,\text{ }k\in \left[ 1,{{k}_{\operatorname{IAC}}} \right] \\
$ $$ $$ $$ $$ $$ $$ $$ $$ $$ $$ $$ $$ $$ $$ $$ $$ $$ $$ $$ $$ $$ $$ $$ $$ $\sum\limits_{k={{k}_{\operatorname{IAC}}}+1}^{K}{{{d}_{k}}}\le M \\
{{d}_{1}}+\sum\limits_{k=1}^{{{k}_{\operatorname{IAC}}}}{(k-1){{d}_{k}}+}\sum\limits_{k={{k}_{\operatorname{IAC}}}+1}^{K}{( {{k}_{\operatorname{IAC}}}-1 ){{d}_{k}}}\le {{k}_{\operatorname{IAC}}}M $ $$ $$ $$ $$ $
\end{IEEEeqnarray}

\section{Design of a Closed-form IAC Transceiver}
In the given example in Fig. 2 and Fig. 3, a solvable set of alignment equations $\Psi$ is available for an illustrative purpose and the corresponding IAC graph automatically satisfies Proposition 1. For any given tuple of DoFs $\left( {{d}_{1}},{{d}_{2}},\cdots,{{d}_{K}} \right)$ that satisfies Theorem 1, however, the set of alignment equations, $\Psi$, is unknown. Different from the classical IA in which a set of alignment equations is unique for a given tuple of DoFs, our STS alignment scheme for IAC provides many different possible $\Psi$’s even for the same tuple of DoFs. This is due to the fact that there are many different possibilities of selecting the basis set $\mathcal{\bar{I}}_{k}^{\operatorname{SIA}}$ and forming the alignment equations in (5). Furthermore, not all of the possible $\Psi$’s are solvable. Therefore, our objective is to find solvable $\Psi$ for any feasible tuple of DoFs.

Meanwhile, ${{k}_{\operatorname{IAC}}}$ can be determined for the given DoFs of $\left( {{d}_{1}},{{d}_{2}},\cdots,{{d}_{K}} \right)$. Then, the associated system overhead ${{O}_{\operatorname{sys}}}$ is fixed, regardless of $\Psi$, by the definition of ${{O}_{\operatorname{sys}}}$. There will be no difference among all possible $\Psi$’s in terms of the total achievable DoFs and system overhead. Hence, we only resort to designing one possible $\Psi$ where the closed-form transmitters $\left\{ {{\mathbf{V}}_{j}} \right\}$ can be solved.

In this section, we establish a framework to design the closed-form transceiver that is addressed by [5]. In Subsection A, we present a design procedure to construct an IAC graph that gives our closed-form transceiver. In Subsection B, we detail how to determine the closed-from transceiver from the constructed IAC graph.

\subsection{Construction of IAC graph for a transceiver design}
Simply due to the enormous complexity involved with the different choices of selecting the basis set and forming the alignment equation, it is not efficient to check the solvability for all possible $\Psi $’s. Considering that $\Psi$ and $\mathcal{G}$ have one-to-one correspondence, we will design a solvable $\Psi$ by constructing a proper IAC graph $\mathcal{G}$. Following the Proposition 1, a constraint on $\mathcal{G}$ for the existence of solutions for $\left\{ {{\mathbf{V}}_{j}} \right\}$ is that each subgraph ${{\mathcal{G}}_{q}}$ does not form more than one loop. Besides, since the interference coming from the same transmitter cannot be aligned at any receiver as dictated in (5), the other constraint on $\mathcal{G}$ is that any two vertices corresponding to the same transmitter cannot be directly connected by one edge or indirectly connected through several edges with the same label.

Focusing on the main flow of the design process, we first elaborate the underlying assumptions and definitions. Considering that the decoding operation in IAC is performed in order, we assume that a design process on $\mathcal{G}$ also follows the same order, namely, from receiver $1$ to receiver $K$. Initially, suppose that there exist only vertices, i.e., without any edge in $\mathcal{G}$, which is denoted by a graph ${{\mathcal{G}}^{\left[ 0 \right]}}=\left( {{\mathcal{P}}^{\left[ 0 \right]}},{{\mathcal{E}}^{\left[ 0 \right]}} \right)$, where ${{\mathcal{E}}^{\left[ 0 \right]}}=\phi$. As a total number of ${{d}_{1}}$ desired symbols for the 1st Tx/Rx pair will be directly cancelled by IC operation in the remaining receiver, $\left\{ {{\mathbf{v}}_{1,{{\ell }_{1}}}} \right\}_{{{\ell }_{1}}=1}^{{{d}_{1}}}$ do not appear in $\Psi$ and therefore, ${{\mathcal{P}}^{\left[ 0 \right]}}=\left\{ \left. {{\mathbf{v}}_{j,{{\ell }_{j}}}} \right|\forall j\in \left[ 2,K \right],\text{ }\forall {{\ell }_{j}}\in \left[ 1,{{d}_{j}} \right] \right\}$.  Assume that each vertex in ${{\mathcal{P}}^{\left[ 0 \right]}}$ can be associated with an individual subgraph of ${{\mathcal{G}}^{\left[ 0 \right]}}$, leading to ${{Q}^{\left[ 0 \right]}}=\left| {{\mathcal{P}}^{\left[ 0 \right]}} \right|$ subgraphs, where ${{Q}^{\left[ k \right]}}$ denotes the number of subgraphs in ${{\mathcal{G}}^{\left[ k \right]}}$, $k=0,1,\cdots ,K$. Then, the edges corresponding to the alignment equations from receivers $1$ to $K$ will be added on ${{\mathcal{G}}^{\left[ 0 \right]}}$ sequentially. If one edge is added between one vertex in subgraph $i$ and the other in subgraph $j$ in the process of edge addition, two different subgraphs, $i$ and $j$, would be merged. Let $\mathcal{\overset{\scriptscriptstyle\smile}{G}}=\left( \mathcal{\overset{\scriptscriptstyle\smile}{P}},\mathcal{\overset{\scriptscriptstyle\smile}{E}} \right)$ and ${{\mathcal{G}}^{\left[ k \right]}}=\left( {{\mathcal{P}}^{\left[ k \right]}},{{\mathcal{E}}^{\left[ k \right]}} \right)$ denote a temporary IAC graph in the edge-addition process and one after the edge-addition operation for receiver $k$, respectively. Therefore, we have $\left| {{\mathcal{P}}^{\left[ k \right]}} \right|=\left| {{\mathcal{P}}^{\left[ 0 \right]}} \right|$ and $\left| {{\mathcal{E}}^{\left[ k \right]}} \right|=\sum\nolimits_{i=1}^{k}{\left| {{\Phi }_{i}} \right|}$ for $k=1,2,\cdots ,K$. Finally, IAC graph is given by $\mathcal{G}={{\mathcal{G}}^{\left[ K \right]}}$. Since the interferences received at receivers $\left( {{k}_{\operatorname{IAC}}}+1 \right)$ to $K$ do not require to be aligned, however, no edges are added for $\left\{ {{\mathcal{G}}^{\left[ k \right]}} \right\}_{k={{k}_{\operatorname{IAC}}}}^{K}$ and thus, $\mathcal{G}={{\mathcal{G}}^{\left[ {{k}_{\operatorname{IAC}}} \right]}}$ and $Q={{Q}^{\left[ {{k}_{\operatorname{IAC}}} \right]}}$.

Let ${{\eta }_{k}}$ and ${{R}_{k}}$ denote a set of vertices for the precoding vectors in interference vectors of $\mathcal{I}_{k}^{\operatorname{IA}}$ and $\mathcal{\bar{I}}_{k}^{\operatorname{SIA}}$ at receiver $k$, respectively. Furthermore, let ${{{\eta }'}_{k}}$ denote a set of vertices for the precoding vectors in $\left( \mathcal{I}_{k}^{\operatorname{IA}}-\mathcal{\bar{I}}_{k}^{\operatorname{SIA}} \right)$, i.e., ${{{\eta }'}_{k}}={{\eta }_{k}}-~{{R}_{k}}$. Furthermore, let $A_{j}^{\left[ k \right]}$ denote a set of reference vertices to which the vertices $\left\{ {{\mathbf{v}}_{j{{\ell }_{j}}}} \right\}_{{{\ell }_{j}}=1}^{{{d}_{j}}}$ have been connected, and initially we have $A_{j}^{\left[ k \right]}={{R}_{k}}\bigcap$ $\left\{ {{\mathbf{v}}_{j{{\ell }_{j}}}} \right\}_{{{\ell }_{j}}=1}^{{{d}_{j}}}$. The following procedure details how $\mathcal{G}$ is designed so as to satisfy two aforementioned constraints:

\textbf{Step 1 (Initialization):} List the total number of $\left| {{\mathcal{P}}^{\left[ 0 \right]}} \right|=\sum\nolimits_{j=2}^{K}{{{d}_{j}}}$ vertices which represent $\left\{ \left. {{\mathbf{v}}_{j{{\ell }_{j}}}} \right|\forall j\in \left[ 2,K \right],\text{} \forall {{\ell }_{j}}\in \left[ 1,{{d}_{j}} \right] \right\}$ and obtain ${{\mathcal{G}}^{\left[ 0 \right]}}=\left( {{\mathcal{P}}^{\left[ 0 \right]}},{{\mathcal{E}}^{\left[ 0 \right]}} \right)$ where ${{\mathcal{P}}^{\left[ 0 \right]}}=\left\{ \left. {{\mathbf{v}}_{j,{{\ell }_{j}}}} \right|\forall j\in \left[ 2,K \right],\text{ }\forall {{\ell }_{j}}\in \left[ 1,{{d}_{j}} \right] \right\}$ and ${{\mathcal{E}}^{\left[ 0 \right]}}=\phi $.

\textbf{Step 2 (Selecting a set of reference vertices):} At the $k$-th Tx/Rx pair, let ${{\eta }_{k}}=\left\{ \left. {{\mathbf{v}}_{{k}'{{\ell }_{{{k}'}}}}} \right|\forall {k}'\in \left[ k+1,K \right],\text{ }\forall {{\ell }_{{{k}'}}}\in \left[ 1,{{d}_{{{k}'}}} \right] \right\}$. Randomly pick up a total of $\left| \mathcal{I}_{k}^{\operatorname{IA}} \right|=M-{{d}_{k}}$ vertices from ${{\eta }_{k}}$ so as to form ${{R}_{k}}$ such that $\left| {{R}_{k}} \right|=M-{{d}_{k}}$. Initially, set ${{\mathcal{G}}^{\left[ k \right]}}\leftarrow {{\mathcal{G}}^{\left[ k-1 \right]}}$.

\textbf{Step 3 (Choosing the alignment pairs):} If $P_{q}^{\left[ k \right]}\bigcap {{{\eta }'}_{k}}\ne \phi $, pick up one vertex ${{\mathbf{v}}_{{j}'{{\ell }_{{{j}'}}}}}\in \mathcal{P}_{q}^{\left[ k \right]}\bigcap {{{\eta }'}_{k}}$. If ${{R}_{k}}-\left( A_{{{j}'}}^{\left[ k \right]}\bigcap {{R}_{k}} \right)=\phi $, go back to Step 2; otherwise, connect ${{\mathbf{v}}_{{j}'{{\ell }_{{{j}'}}}}}$ to one reference vertex ${{\mathbf{v}}_{j{{\ell }_{j}}}}\in {{R}_{k}}-\left( A_{{{j}'}}^{\left[ k \right]}\bigcap {{R}_{k}} \right)$ by an edge $e_{\left( j\ell  \right),\left( j'{\ell }' \right)}^{\left[ k \right]}$ which represents the alignment equation of $\operatorname{span}\left( {{\mathbf{H}}_{kj}}{{\mathbf{v}}_{j\ell }} \right)=\operatorname{span}\left( {{\mathbf{H}}_{kj'}}{{\mathbf{v}}_{j'{\ell }'}} \right)$. Then, we have 
\[\mathcal{\overset{\scriptscriptstyle\smile}{E}}_{q}^{{}}=\left\{ \begin{array}{*{35}{l}}
   \mathcal{E}_{q}^{\left[ k \right]}\bigcup \left\{ e_{\left( j\ell  \right),\left( j'{\ell }' \right)}^{\left[ k \right]} \right\}, & {{\mathbf{v}}_{j{{\ell }_{j}}}}\in \mathcal{P}_{q}^{\left[ k \right]}  \\
   \mathcal{E}_{q}^{\left[ k \right]}\bigcup \mathcal{E}_{{{q}'}}^{\left[ k \right]}\bigcup \left\{ e_{\left( j\ell  \right),\left( j'{\ell }' \right)}^{\left[ k \right]} \right\}, & {{\mathbf{v}}_{j{{\ell }_{j}}}}\in \mathcal{P}_{{{q}'}}^{\left[ k \right]},\text{ }{q}'\ne q  \\
\end{array} \right.\]
and 
\[\mathcal{\overset{\scriptscriptstyle\smile}{P}}_{q}^{{}}=\left\{ \begin{array}{*{35}{l}}
   \mathcal{P}_{q}^{\left[ k \right]}, & {{\mathbf{v}}_{j{{\ell }_{j}}}}\in \mathcal{P}_{q}^{\left[ k \right]}  \\
   \mathcal{P}_{q}^{\left[ k \right]}\bigcup \mathcal{P}_{{{q}'}}^{\left[ k \right]}, & {{\mathbf{v}}_{j{{\ell }_{j}}}}\in \mathcal{P}_{{{q}'}}^{\left[ k \right]},\text{ }{q}'\ne q  \\
\end{array} \right.\]

\textbf{Step 4 (Checking the number of formed loops):} If $\left| \mathcal{\overset{\scriptscriptstyle\smile}{E}}_{q}^{{}} \right|\le \left| \mathcal{\overset{\scriptscriptstyle\smile}{P}}_{q}^{{}} \right|$,  update ${{\mathcal{G}}^{\left[ k \right]}}$ by $\mathcal{E}_{q}^{\left[ k \right]}=\mathcal{\overset{\scriptscriptstyle\smile}{E}}_{q}^{{}}$, $\mathcal{P}_{q}^{\left[ k \right]}=\mathcal{\overset{\scriptscriptstyle\smile}{P}}_{q}^{{}}$, $\mathcal{E}_{{{q}'}}^{\left[ k \right]}\leftarrow \left\{ \begin{array}{*{35}{l}}
   \mathcal{E}_{{{q}'}}^{\left[ k \right]}, & {{\mathbf{v}}_{j{{\ell }_{j}}}}\in \mathcal{P}_{q}^{\left[ k \right]}  \\
   \phi , & {{\mathbf{v}}_{j{{\ell }_{j}}}}\in \mathcal{P}_{{{q}'}}^{\left[ k \right]},\text{ }{q}'\ne q  \\
\end{array} \right.$, and $\mathcal{P}_{{{q}'}}^{\left[ k \right]}\leftarrow \left\{ \begin{array}{*{35}{l}}
   \mathcal{P}_{{{q}'}}^{\left[ k \right]}, & {{\mathbf{v}}_{j{{\ell }_{j}}}}\in \mathcal{P}_{q}^{\left[ k \right]}  \\
   \phi , & {{\mathbf{v}}_{j{{\ell }_{j}}}}\in \mathcal{P}_{{{q}'}}^{\left[ k \right]},\text{ }{q}'\ne q  \\
\end{array} \right.$,
\\ while updated as $A_{{{j}'}}^{\left[ k \right]}\leftarrow A_{{{j}'}}^{\left[ k \right]}\bigcup \left\{ {{\mathbf{v}}_{j{{\ell }_{j}}}} \right\}$ and ${{{\eta }'}_{k}}\leftarrow {{{\eta }'}_{k}}-\{ {{\mathbf{v}}_{{j}'{{\ell }_{{{j}'}}}}} \}$; if $\left| \mathcal{\overset{\scriptscriptstyle\smile}{E}}_{q}^{{}} \right|>\left| \mathcal{\overset{\scriptscriptstyle\smile}{P}}_{q}^{{}} \right|$, go back to Step 3 to change another reference vertex for ${{\mathbf{v}}_{{j}'{{\ell }_{{{j}'}}}}}$.

\textbf{Step 5:} Repeat Steps 3 and 4 until $\mathcal{P}_{q}^{\left[ k \right]}\bigcap {{{\eta }'}_{k}}=\phi$.

\textbf{Step 6:} Repeat Steps 3 to 5 for $q\in \left[ 1,{{Q}^{\left[ k \right]}} \right]$ and then, ${{\mathcal{G}}^{\left[ k \right]}}$ is obtained.

\textbf{Step 7:} Repeat Steps 2 to 6 for $k\in \left[ 2,{{k}_{\operatorname{IAC}}} \right]$ and finally, we have $\mathcal{G}={{\mathcal{G}}^{\left[ {{k}_{\operatorname{IAC}}} \right]}}$.

In Step 3, by choosing a reference vertex ${{\mathbf{v}}_{j{{\ell }_{j}}}}\in {{R}_{k}}-\left( A_{{{j}'}}^{\left[ k \right]}\bigcap {{R}_{k}} \right)$, it guarantees that two vertices from the same transmitter ${j}'$ are not connected by one edge or indirectly connected through several edges with the same label.

In the sequel, a simple proof can be sketched to show that the proposed design has formed at most one loop in each subgraph ${{\mathcal{G}}_{q}}$. At the 1st Tx/Rx pair, following Steps 2 to 6, a total of $\left| {{R}_{1}} \right|=M-{{d}_{1}}$ reference vertices are selected. As each vertex ${{\mathbf{v}}_{{j}'{{\ell }_{{{j}'}}}}}\in {{{\eta }'}_{1}}$ can be connected to only one unique reference vertex and no edge is allowed between any two reference vertices, a total of ${{Q}^{\left[ 1 \right]}}=M-{{d}_{1}}$ subgraphs have been formed in ${{\mathcal{G}}^{\left[ 1 \right]}}$ with $\left| \mathcal{E}_{q}^{\left[ 1 \right]} \right|=\left| \mathcal{P}_{q}^{\left[ 1 \right]} \right|-1$, $q\in \left[ 1,{{Q}^{\left[ 1 \right]}} \right]$. Then, the alignment equations from 2nd to ${{k}_{\operatorname{IAC}}}$-th Tx/Rx pair are added onto ${{\mathcal{G}}^{\left[ 1 \right]}}$, giving a total number of $( \sum\nolimits_{k=2}^{{{k}_{\operatorname{IAC}}}}{\left| \Phi {}_{k} \right|}=\sum\nolimits_{k=2}^{{{k}_{\operatorname{IAC}}}}{\left( k-1 \right){{d}_{k}}}+\sum\nolimits_{{k}'={{k}_{\operatorname{IAC}}}+1}^{K}{\left( {{k}_{\operatorname{IAC}}}-1 \right){{d}_{{{k}'}}}}-\left( {{k}_{\operatorname{IAC}}}-1 \right)M )$ edges. Referring to Theorem 1, we can have $\sum\nolimits_{k=2}^{{{k}_{\operatorname{IAC}}}}{\left| \Phi {}_{k} \right|}\le M-{{d}_{1}}$, which indicates that the total number of edges added on ${{\mathcal{G}}^{\left[ 1 \right]}}$ by the 2nd to ${{k}_{\operatorname{IAC}}}$-th Tx/Rx pairs is no more than the total number of subgraphs in ${{\mathcal{G}}^{\left[ 1 \right]}}$. Furthermore, by going through Step 4, the proposed design guarantees that no more than one edge is added in each subgraph $\mathcal{G}_{q}^{\left[ k \right]}$. According to graph theory, if one additional edge is added on an acyclic subgraph, then one loop would be formed. On the other hand, if one edge is added between two vertices that belong to two different subgraphs, then two subgraphs are merged into one. Consequently, we have shown that the proposed design has constructed IAC graph $G$ to satisfy Proposition 1. In other words, the closed-form transmitters $\{{{\mathbf{V}}_{j}}\}_{j=1}^{K}$ exist and can be solved from $\Psi$.

\subsection{Solution to the closed-form IAC transceivers}
\begin{figure}[!t]
\vspace*{-10pt}
\centering{
\subfloat[${{\mathcal{G}}_{A}}=({{\mathcal{P}}_{A}},{{\mathcal{E}}_{A}})$: No-loop case$ $$ $$ $]{\includegraphics[width=1.7in]{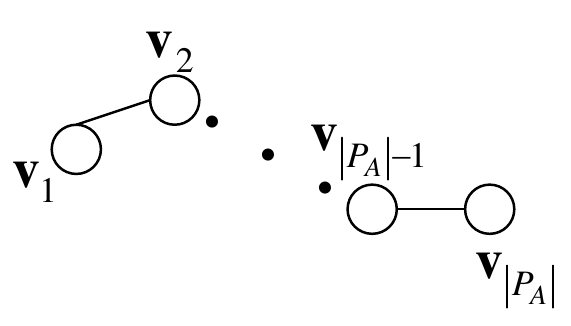}}
\subfloat[${{\mathcal{G}}_{B}}=({{\mathcal{P}}_{B}},{{\mathcal{E}}_{B}})$: One-loop case]{\includegraphics[width=1.7in]{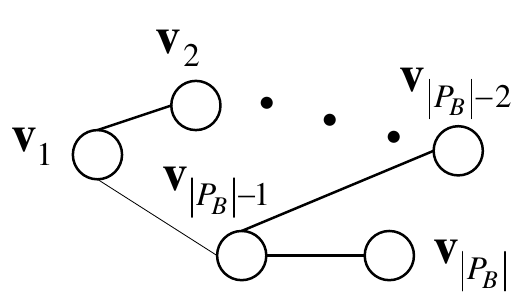}}}
\caption{Two types of subgraphs in IAC graph $\mathcal{G}$: No-loop case vs. one-loop case}
\vspace*{-10pt}
\end{figure}



In the previous subsection, we have presented a general design principle on $\mathcal{G}$ to specify a solvable $\Psi$. In this subsection, we present the procedures to solve the closed-form transceivers through $\mathcal{G}$. For the subgraphs that satisfy Proposition 1, there are two distinct cases: no-loop case (subgraph ${{\mathcal{G}}_{A}}$) and one-loop case, as illustrated in Fig. 4. For a general illustration, without loss of generality, we do not give a specific label to each edge, and constrain each vertex to any specific user for the examples in Fig. 4. For a simplicity of exposition, therefore, every vertex can be denoted by the indices $\left\{ {{\mathbf{v}}_{p}} \right\}_{p=1}^{|\mathcal{P}|}$. Then, a simplified definition of the channel matrices, without clarifying the transmitter and receiver index, will be adopted, i.e., $\{\mathbf{H}_{p}^{\left( 1 \right)}\}_{p=1}^{\left| \mathcal{P} \right|}$, $\{ \mathbf{H}_{p}^{\left( 2 \right)}\}_{p=1}^{\left| \mathcal{P} \right|}$, $\cdots$, and $\{ \mathbf{H}_{p}^{\left( {{L}_{p}} \right)}\}_{p=1}^{\left| \mathcal{P} \right|}$ denote the channel matrices for ${{\mathbf{v}}_{p}}$, while using superscripts $\left( 1 \right)$, $\left( 2 \right)$, $\cdots$, and $\left( {{L}_{p}} \right)$ to distinguish the channel matrices for ${{\mathbf{v}}_{p}}$ in the different alignment equations in (5). Additionally, $\{ \mathbf{H}_{p}^{\left( i \right)}\}$ is equal to $\{ \mathbf{H}_{p}^{\left( j \right)}\}$ for $\forall i,j\in \left\{ 1,2,\cdots ,{{L}_{p}} \right\}$ if the two different equations that are involved with $\{ \mathbf{H}_{p}^{\left( i \right)}\}$ and $\{ \mathbf{H}_{p}^{\left( j \right)}\}$, respectively, belong to the same receiver; otherwise, they are different. The corresponding set of linear alignment equations, ${{\Phi }_{A}}$, that are specified by the subgraph ${{\mathcal{G}}_{A}}=({{\mathcal{P}}_{A}},{{\mathcal{E}}_{A}})$ in Fig. 4(a) is given as
\begin{IEEEeqnarray}{lcr}
\left\{ \begin{matrix}
   \operatorname{span}\left( \mathbf{H}_{1}^{\left( 1 \right)}{{\mathbf{v}}_{1}} \right)=\operatorname{span}\left( \mathbf{H}_{2}^{\left( 1 \right)}{{\mathbf{v}}_{2}} \right)  \\
   \operatorname{span}\left( \mathbf{H}_{2}^{\left( 2 \right)}{{\mathbf{v}}_{2}} \right)=\operatorname{span}\left( \mathbf{H}_{3}^{\left( 1 \right)}{{\mathbf{v}}_{3}} \right)  \\
   \vdots   \\
   \operatorname{span}\left( \mathbf{H}_{\left| {{P}_{A}} \right|-1}^{\left( 2 \right)}{{\mathbf{v}}_{\left| {{P}_{A}} \right|-1}} \right)=\operatorname{span}\left( \mathbf{H}_{\left| {{P}_{A}} \right|}^{\left( 1 \right)}{{\mathbf{v}}_{\left| {{P}_{A}} \right|}} \right)  \\
\end{matrix} \right.  
\end{IEEEeqnarray}      
For any equation of $\operatorname{span}(\mathbf{H}_{{{p}_{1}}}^{\left( i \right)}{{\mathbf{v}}_{{{p}_{1}}}} )=\operatorname{span}( \mathbf{H}_{{{p}_{2}}}^{\left( j \right)}{{\mathbf{v}}_{{{p}_{2}}}})$ in (12), it can be expressed equivalently by ${{\mathbf{v}}_{{{p}_{2}}}}={{\gamma }_{{{p}_{2}},{{p}_{1}}}}{( \mathbf{H}_{{{p}_{2}}}^{\left( j \right)})^{-1}}\mathbf{H}_{{{p}_{1}}}^{\left( i \right)}{{v}_{{{p}_{1}}}}$, where ${{\gamma }_{{{p}_{2}},{{p}_{1}}}}$ is the coefficient that coordinates the length of two vectors, $\mathbf{H}_{{{p}_{1}}}^{\left( i \right)}{{\mathbf{v}}_{{{p}_{1}}}}$ and $\mathbf{H}_{{{p}_{2}}}^{\left( j \right)}{{\mathbf{v}}_{{{p}_{2}}}}$, and would be determined while solving the equations. As $\{{{\mathbf{v}}_{j}}\}_{j=2}^{_{\left| {\mathcal{{P}}_{A}} \right|}}$ can be expressed in terms of ${{\mathbf{v}}_{1}}$ in (12), ${{\mathbf{v}}_{1}}$ can be selected as any arbitrary vector, i.e., assume that ${{\mathbf{v}}_{1}}$ is an $M\times 1$ unit vector and then, $\{{{\mathbf{v}}_{j}}\}_{j=2}^{_{\left| {\mathcal{{P}}_{A}} \right|}}$ can be solved sequentially. 

The following set of linear alignment equations ${{\Phi }_{B}}$ is given by the subgraph ${{\mathcal{G}}_{B}}=({{\mathcal{P}}_{B}},{{\mathcal{E}}_{B}})$ in Fig. 4(b):
\begin{IEEEeqnarray}{lcr}
\left\{ \begin{matrix}
   \operatorname{span}( \mathbf{H}_{1}^{\left( 1 \right)}{{\mathbf{v}}_{1}} )=\operatorname{span}( \mathbf{H}_{2}^{\left( 1 \right)}{{\mathbf{v}}_{2}} )  \\
   \vdots   \\
   \operatorname{span}( \mathbf{H}_{\left| {\mathcal{{P}}_{B}} \right|-2}^{\left( 2 \right)}{{\mathbf{v}}_{\left| {\mathcal{{P}}_{B}} \right|-2}} )=\operatorname{span}( \mathbf{H}_{\left| {\mathcal{{P}}_{B}} \right|-1}^{\left( 1 \right)}{{\mathbf{v}}_{\left| {\mathcal{{P}}_{B}} \right|-1}} )  \\
   \operatorname{span}( \mathbf{H}_{\left| {\mathcal{{P}}_{B}} \right|-1}^{\left( 2 \right)}{{\mathbf{v}}_{\left| {\mathcal{{P}}_{B}} \right|-1}} )=\operatorname{span}( \mathbf{H}_{1}^{\left( 2 \right)}{{\mathbf{v}}_{1}} )  \\
   \operatorname{span}( \mathbf{H}_{\left| {\mathcal{{P}}_{B}} \right|-1}^{\left( 3 \right)}{{\mathbf{v}}_{\left| {\mathcal{{P}}_{B}} \right|-1}} )=\operatorname{span}( \mathbf{H}_{\left| {\mathcal{{P}}_{B}} \right|}^{\left( 1 \right)}{{\mathbf{v}}_{\left| {\mathcal{{P}}_{B}} \right|}} )
\end{matrix} \right.                    
\end{IEEEeqnarray}
Considering the first $(\left| {{\mathcal{P}}_{B}} \right|-1)$ equations in (13), which are obtained around the loop in ${{\mathcal{G}}_{B}}$, it can be shown that ${{\mathbf{v}}_{1}}={{\mathbf{F}}_{1}}{{\mathbf{v}}_{1}}$, i.e., ${{\mathbf{v}}_{1}}$ is the eigenvector of ${{\mathbf{F}}_{1}}$, where ${{\mathbf{F}}_{1}}={{\gamma }_{1,\left| {{\mathcal{P}}_{B}} \right|-1}}{( \mathbf{H}_{1}^{\left( 2 \right)})^{-1}}\left( \prod\nolimits_{j=\left| {{\mathcal{P}}_{B}} \right|-1}^{2}{{{\gamma }_{j,j-1}}\mathbf{H}_{j}^{\left( 2 \right)}( \mathbf{H}_{j}^{\left( 1 \right)})^{-1}} \right)\mathbf{H}_{1}^{\left( 1 \right)}$.\footnotemark \\ For the other precoding vectors $\{{{\mathbf{v}}_{j}}\}_{j=2}^{|{\mathcal{{P}}_{B}}|}$, they could be expressed in terms of ${{\mathbf{v}}_{1}}$ by (13) as we did in the no-loop case. Therefore, once ${{\mathbf{v}}_{1}}$ is known, $\{{{\mathbf{v}}_{j}}\}_{j=2}^{|{\mathcal{{P}}_{B}}|}$ can be solved sequentially. In summary, the closed-form solutions of $\left\{ \left. {{\mathbf{v}}_{j{{\ell }_{j}}}} \right|\forall {{\ell }_{j}}\in \left[ 1,{{d}_{j}} \right],\text{ }\forall j\in \left[ 2,K \right] \right\}$ can be found through two ways as discussed in the above, depending on the type of the involved subgraph. Furthermore, as $\{{{\mathbf{v}}_{1{{\ell }_{1}}}}\}_{{{\ell }_{1}}=1}^{{{d}_{1}}}$ do not appear in $\mathcal{G}$, they can be immediately found by ${{\mathbf{V}}_{1}}\subset {{\left( {{\mathbf{H}}_{1j}} \right)}^{-1}}{{\left( \mathcal{I}_{1}^{\operatorname{IA}} \right)}^{\bot }}$ where ${{\mathcal{I}}_{1}}$ is known from $\left\{ \left. {{\mathbf{v}}_{j{{\ell }_{j}}}} \right|\forall {{\ell }_{j}}\in \left[ 1,{{d}_{j}} \right],\text{ }\forall j\in \left[ 2,K \right] \right\}$. Once the transmitters $\{{{\mathbf{V}}_{j}}\}_{j=1}^{K}$ are solved, $\{\mathcal{I}_{k}^{\operatorname{IA}}\}_{k=1}^{K}$ is determined and then, the receiver ${{\mathbf{U}}_{k}}$ can be obtained by ${{\mathbf{U}}_{k}}\subset {{\left( \mathcal{I}_{k}^{\operatorname{IA}} \right)}^{\bot }}$, $\forall k\in \left[ 1,K \right]$. In the following subsection, we present an illustrative example for the current design principle to solve a closed-form IAC transceiver that can achieve the maximum degree of freedom.
\footnotetext{The indices of product operation “$\prod\nolimits_{j=\left| {{\mathcal{P}}_{B}} \right|-1}^{2}$” incrementally decreasing from the lower limit to the upper limit, indicate the order for the matrix multiplication, i.e., the matrix with the bigger index is post-multiplied by one with the smaller one.}

\subsection{Illustrative example}
In the last subsection, we have presented the design principle for a closed-form transceiver for any feasible tuple of DoFs. Naturally, among all feasible tuples of DoFs with the same $K$ and $M$, we will be interested in finding the optimum tuple which maximizes the total achievable DoFs. In our previous work [6], we have shown that the total achievable DoF by the closed-form IAC solutions is given by $2M$. Moreover, we have also shown that it can be achieved by ${{k}_{\operatorname{IAC}}}=2$. In this subsection, we consider an illustrative example for the proposed design procedure to solve a closed-form IAC transceiver that can achieve the maximum degree of freedom. In other words, our IAC transceiver must be designed to meet the constraint on the tuple of DoFs $\left( {{d}_{1}},{{d}_{2}},\cdots ,{{d}_{K}} \right)$ which can achieve a total of $2M$ for ${{k}_{\operatorname{IAC}}}=2$. In other words, our DoF constraint must follow (9) for ${{k}_{\operatorname{IAC}}}=2$, i.e., 
\begin{IEEEeqnarray}{lcr}
\left\{ \begin{matrix}
   {{d}_{1}}+\max \left\{ {{d}_{j}} \right\}_{j=2}^{K}\le M  \\
   {{d}_{2}}+\max \left\{ {{d}_{j}} \right\}_{j=3}^{K}\le M  
\end{matrix} \right.
\end{IEEEeqnarray}  
Then, (10) and (11) reduces to be  $\sum\nolimits_{k=3}^{K}{{{d}_{k}}}\le M$ and ${{d}_{1}}+{{d}_{2}}+\sum\nolimits_{k=3}^{K}{{{d}_{k}}}\le 2M$, respectively. Let $\sum\nolimits_{k=3}^{K}{{{d}_{k}}}=M$ and ${{d}_{1}}+{{d}_{2}}=M$. Then, a total of $2M$ can be collected. Substituting ${{d}_{1}}+{{d}_{2}}=M$ into (14) and combining the two inequalities, we can get
\begin{IEEEeqnarray}{lcr}
\max \left\{ {{d}_{j}} \right\}_{j=3}^{K}\le \frac{M}{2}
\end{IEEEeqnarray}
Therefore, the constraints on $\left( {{d}_{1}},{{d}_{2}},\cdots ,{{d}_{K}} \right)$ can be summarized as
\begin{IEEEeqnarray}{lcr} 
\left\{ \begin{array}{*{35}{l}}
   {{d}_{1}}\le M-\max \left\{ {{d}_{j}} \right\}_{j=3}^{K}  \\
   {{d}_{2}}\le M-\max \left\{ {{d}_{j}} \right\}_{j=3}^{K}  \\
   \max \left\{ {{d}_{j}} \right\}_{j=3}^{K}\le \frac{M}{2}  \\
   {{d}_{1}}+{{d}_{2}}=M  \\
   \sum\nolimits_{j=3}^{K}{{{d}_{j}}}=M  \\
\end{array} \right.      
\end{IEEEeqnarray}                     
From the third and fifth inequalities (16), we can get $K\ge 4$.

\begin{figure}[!t]
\centering
\includegraphics[width=3.4in]{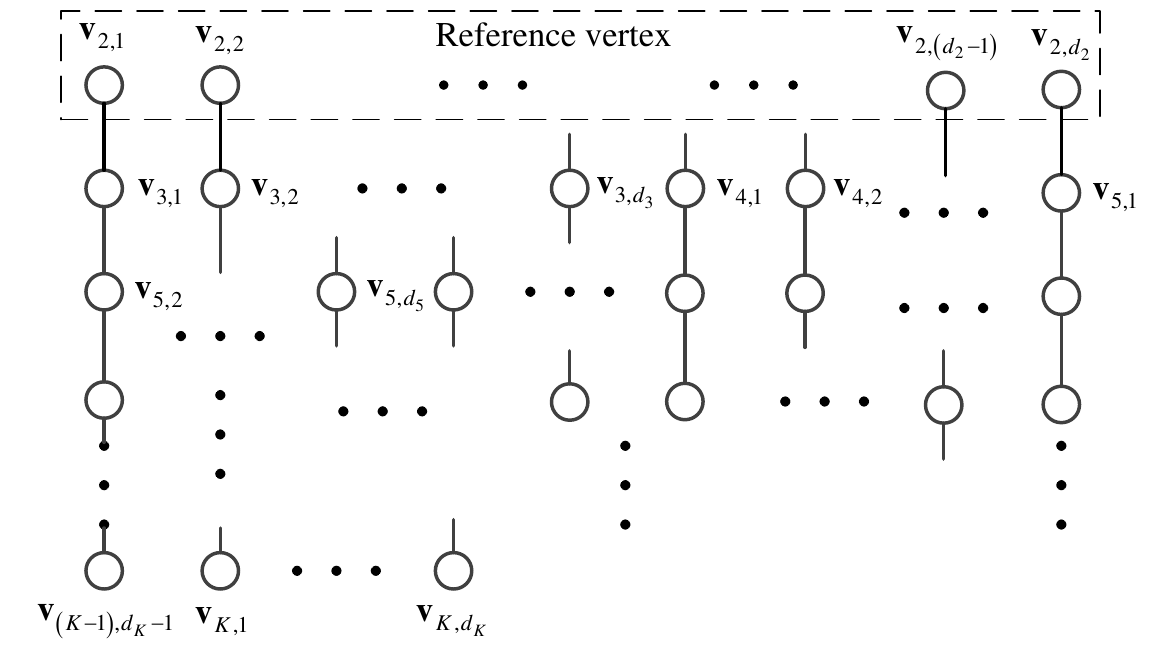}
\vspace*{-10pt}
\caption{Optimal transceiver design with ${{k}_{\operatorname{IAC}}}=2$: ${{\mathcal{G}}^{\left[ 1 \right]}}=\left( {{\mathcal{P}}^{\left[ 1 \right]}},{{\mathcal{E}}^{\left[ 1 \right]}} \right)$}
\end{figure}

In the sequel, we consider a design of the closed-form transceiver with $\left( {{d}_{1}},{{d}_{2}},\cdots ,{{d}_{K}} \right)$ which satisfies (16). Following the general design principle in the previous subsection, we design a solvable $\Psi $ by constructing an IAC graph $\mathcal{G}$. With ${{k}_{\operatorname{IAC}}}=2$, we have $\mathcal{G}={{\mathcal{G}}^{\left[ 2 \right]}}$ and therefore, we only resort to ${{\mathcal{G}}^{\left[ 1 \right]}}$ and ${{\mathcal{G}}^{\left[ 2 \right]}}$ to obtain $\mathcal{G}$. As opposed to the example in Fig. 2, which achieves a total DoFs of $11$, our target transceiver is supposed to achieve the maximum total DoFs, $2M=12$, with the same system structure of $K=5$ and $M=6$. Referring to the constraints in (16), we consider a system of ${{d}_{1}}={{d}_{2}}=3$ and ${{d}_{3}}={{d}_{4}}={{d}_{5}}=2$, which achieves a total DoF of $12$, as an illustrative design example.  

We first provide the design of ${{\mathcal{G}}^{\left[ 1 \right]}}$. At receiver 1, we have ${{\eta }_{1}}=\left\{ \left. {{\mathbf{v}}_{{j}'{{\ell }_{{{j}'}}}}} \right|\forall {j}'\in \left[ 2,K \right],\text{ }\forall {{\ell }_{{{j}'}}}\in \left[ 1,{{d}_{{{j}'}}} \right] \right\}$. Considering that $\left| \mathcal{\bar{I}}_{1}^{\operatorname{SIA}} \right|=M-{{d}_{1}}={{d}_{2}}$, it inspires us to construct a set of the reference vertices as ${{R}_{1}}=\left\{ \left. {{\mathbf{v}}_{2{{\ell }_{2}}}} \right|\forall {{\ell }_{2}}\in \left[ 1,{{d}_{2}} \right] \right\}$ and then, the remaining vertices are given as ${{{\eta }'}_{1}}={{\eta }_{1}}-{{R}_{1}}=\left\{ \left. {{\mathbf{v}}_{{j}'{{\ell }_{{{j}'}}}}} \right|\forall {j}'\in \left[ 3,K \right],\text{ }\forall {{\ell }_{{{j}'}}}\in \left[ 1,{{d}_{{{j}'}}} \right] \right\}$. Initially, set ${{\mathcal{G}}^{\left[ 1 \right]}}\leftarrow {{\mathcal{G}}^{\left[ 0 \right]}}$. Suppose a total of ${{d}_{2}}$ reference vertices are artificially arranged by the index ${{\ell }_{2}}$ as shown in Fig. 5. Firstly, each vertex ${{\mathbf{v}}_{3{{\ell }_{3}}}}$, $\forall {{\ell }_{3}}\in \left[ 1,{{d}_{3}} \right]$, will be connected to one unique reference vertex ${{\mathbf{v}}_{2{{\ell }_{2}}}}$, $\forall {{\ell }_{2}}\in \left[ 1,{{d}_{2}} \right]$, with ${{\ell }_{2}}={{\ell }_{3}}\bmod \left( {{d}_{2}} \right)$, creating an edge $e_{\left( 2{{\ell }_{2}} \right),\left( 3{{\ell }_{3}} \right)}^{\left[ 1 \right]}$ which represents an alignment equation, given by  $\operatorname{span}\left( {{\mathbf{H}}_{13}}{{\mathbf{v}}_{3{{\ell }_{3}}}} \right)=\operatorname{span}\left( {{\mathbf{H}}_{12}}{{\mathbf{v}}_{2{{\ell }_{2}}}} \right)$, and then, we have $\mathcal{E}_{{{\ell }_{2}}}^{\left[ 1 \right]}=\mathcal{E}_{{{\ell }_{2}}}^{\left[ 1 \right]}\bigcup \left\{ e_{\left( 2{{\ell }_{2}} \right),\left( 3{{\ell }_{3}} \right)}^{\left[ 1 \right]} \right\}$. After finishing the connection of all vertices $\left\{ {{\mathbf{v}}_{3{{\ell }_{3}}}} \right\}_{{{\ell }_{3}}=1}^{{{d}_{3}}}$ from the 3rd Tx, we turn to connect the ones from the 4th Tx and so as the other Tx’s. Finally, we obtain ${{\mathcal{G}}^{\left[ 1 \right]}}=\left( {{\mathcal{P}}^{\left[ 1 \right]}},{{\mathcal{E}}^{\left[ 1 \right]}} \right)$ with $\left| {{\mathcal{P}}^{\left[ 1 \right]}} \right|=\sum\nolimits_{j=2}^{K}{{{d}_{j}}}={{d}_{2}}+M$ and $\left| {{\mathcal{E}}^{\left[ 1 \right]}} \right|=\left| \mathcal{I}_{1}^{\operatorname{IA}}\mathcal-{\bar{I}}_{1}^{\operatorname{SIA}} \right|=M$. Furthermore, a corresponding set of linear alignment equations ${{\Phi }_{1}}$, which is specified by ${{\mathcal{G}}^{\left[ 1 \right]}}$, can be expressed as
\begin{IEEEeqnarray}{lcr}
\operatorname{span}\left( {{\mathbf{H}}_{1{j}'}}{{\mathbf{v}}_{{j}'{{\ell }_{{{j}'}}}}} \right)=\operatorname{span}\left( {{\mathbf{H}}_{12}}{{\mathbf{v}}_{2{{\ell }_{2}}}} \right),\text{ }\forall {j}'\in \left[ 3,K \right] $ $$ $            
\end{IEEEeqnarray}
where ${{\ell }_{2}}=\left\{ \begin{array}{*{35}{l}}
   {{\ell }_{{{j}'}}}\bmod \left( {{d}_{2}} \right),\text{ }{j}'=3 
\\ \left( \sum\nolimits_{i=3}^{{j}'-1}{{{d}_{i}}}+{{\ell }_{{{j}'}}} \right)\bmod \left( {{d}_{2}} \right),\text{ }{j}'\ge 4  
\end{array} \right.$.

Since no edge is allowed between any two reference vertices, a total of ${{Q}^{\left[ 1 \right]}}=\left| {{R}_{1}} \right|={{d}_{2}}$ subgraphs have been formed in ${{\mathcal{G}}^{\left[ 1 \right]}}$ after the connection operation of vertices between ${{R}_{1}}$ and ${{{\eta }'}_{1}}$, as shown in Fig. 5. With $\left| {{R}_{1}} \right|={{d}_{2}}$ and $\left| {{{{\eta }'}}_{1}} \right|=M$, the design ensures that at least one vertex has been connected to each reference vertex and thus, there exist at least two vertices in each subgraph ${{\ell }_{2}}$, i.e., $\left| \mathcal{P}_{{{\ell }_{2}}}^{\left[ 1 \right]} \right|\ge 2$. Furthermore, referring to the first and fifth equations in (16), we have $\max \left\{ {{d}_{{{j}'}}} \right\}_{{j}'=3}^{K}\le {{d}_{2}}$, so as to ensure that the vertices from the same Tx are avoided being connected to the same reference vertex. For the illustrative example, ${{\mathcal{G}}^{\left[ 1 \right]}}=\left( {{\mathcal{P}}^{\left[ 1 \right]}},{{\mathcal{E}}^{\left[ 1 \right]}} \right)$ can be obtained as given in Fig. 6.

\begin{figure}[!t]
\centering
\includegraphics[width=1.76in]{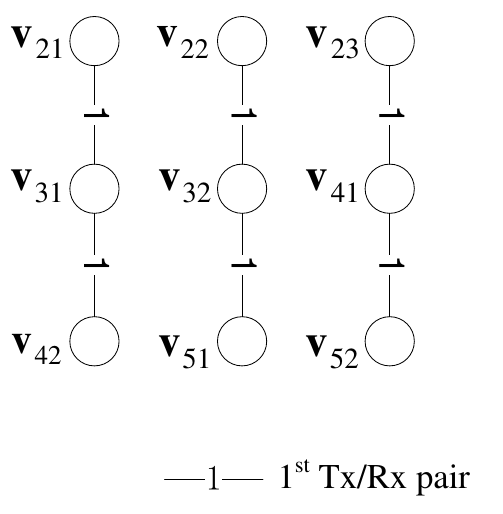}
\vspace*{-10pt}
\caption{Optimal transceiver design of the illustrative example: ${{\mathcal{G}}^{\left[ 1 \right]}}=\left( {{\mathcal{P}}^{\left[ 1 \right]}},{{\mathcal{E}}^{\left[ 1 \right]}} \right)$}
\end{figure}

In the sequel, we provide the design of ${{\mathcal{G}}^{\left[ 2 \right]}}$. Initially, set ${{\mathcal{G}}^{\left[ 2 \right]}}\leftarrow {{\mathcal{G}}^{\left[ 1 \right]}}$. At receiver 2, we have ${{\eta }_{2}}=\{ \left. {{\mathbf{v}}_{{j}'{{\ell }_{{{j}'}}}}} \right|\forall {j}'\in \left[ 3,K \right],\text{ }\forall {{\ell }_{{{j}'}}}\in \left[ 1,{{d}_{{{j}'}}} \right] \}$. With $\left| {{{{\eta }'}}_{2}} \right|=\left| \mathcal{I}_{2}^{\operatorname{IA}}-\mathcal{\bar{I}}_{2}^{\operatorname{SIA}} \right|=\left( \sum\nolimits_{j=3}^{K}{{{d}_{j}}} \right)-\left( M-{{d}_{2}} \right)={{d}_{2}}$, we can form ${{{\eta }'}_{2}}$ by selecting one vertex ${{\mathbf{v}}_{{j}'{{\ell }_{{{j}'}}}}}\in \left( P_{{{{{\ell }'}}_{2}}}^{\left[ 2 \right]}\bigcap {{\eta }_{2}} \right)$ from each $\mathcal{P}_{{{{{\ell }'}}_{2}}}^{\left[ 2 \right]}$, ${{{\ell }'}_{2}}\in \left[ 1,{{d}_{2}} \right]$. This indicates that only one vertex in each $\mathcal{P}_{{{{{\ell }'}}_{2}}}^{\left[ 2 \right]}$ is required to be connected to ${{R}_{2}}$. ${{R}_{2}}$ can be obtained by ${{R}_{2}}={{\eta }_{2}}-{{{\eta }'}_{2}}$. For the illustrative example, we have ${{{\eta }'}_{2}}=\left\{ {{\mathbf{v}}_{31}},{{\mathbf{v}}_{32}},{{\mathbf{v}}_{41}} \right\}$ and ${{R}_{2}}=\left\{ {{\mathbf{v}}_{42}},{{\mathbf{v}}_{51}},{{\mathbf{v}}_{52}} \right\}$, as shown in Fig. 7.

The connection between ${{{\eta }'}_{2}}$ to ${{R}_{2}}$ is completed in two steps. In the first step, for the only vertex ${{\mathbf{v}}_{{j}'{{l}_{{{j}'}}}}}\in {{{\eta }'}_{2}}\bigcap P_{{{\ell }_{2}}}^{\left[ 2 \right]}$ in subgraph $\mathcal{G}_{{{\ell }_{2}}}^{\left[ 2 \right]}$, $\forall {{\ell }_{2}}\in \left[ 1,{{d}_{2}} \right]$, if there is $\left( {{R}_{2}}-A_{{{j}'}}^{\left[ 2 \right]} \right)\bigcap P_{{{\ell }_{2}}}^{\left[ 2 \right]}\ne \phi $, then the vertex ${{\mathbf{v}}_{{j}'{{l}_{{{j}'}}}}}$ is connected to one reference vertex ${{\mathbf{v}}_{j{{l}_{j}}}}\in \left( {{R}_{2}}-A_{{{j}'}}^{\left[ 2 \right]} \right)\bigcap P_{{{\ell }_{2}}}^{\left[ 2 \right]}$, creating an edge $e_{\left( j{{l}_{j}} \right),\left( {j}'{{l}_{{{j}'}}} \right)}^{\left[ 2 \right]}$ which represents alignment equation of $\operatorname{span}\left( {{\mathbf{H}}_{2j}}{{\mathbf{v}}_{j{{\ell }_{j}}}} \right)=\operatorname{span}\left( {{\mathbf{H}}_{2{j}'}}{{\mathbf{v}}_{{j}'{{\ell }_{{{j}'}}}}} \right)$. Therefore, $\mathcal{E}_{{{\ell }_{2}}}^{\left[ 2 \right]}\leftarrow \mathcal{E}_{{{\ell }_{2}}}^{\left[ 2 \right]}\bigcup \left\{ e_{\left( j{{l}_{j}} \right),\left( {j}'{{l}_{{{j}'}}} \right)}^{\left[ 2 \right]} \right\}$ and one loop has been formed in $\mathcal{G}_{{{\ell }_{2}}}^{\left[ 2 \right]}$. Now, update $A_{{{j}'}}^{\left[ 2 \right]}$ by $A_{{{j}'}}^{\left[ 2 \right]}\leftarrow A_{{{j}'}}^{\left[ 2 \right]}\bigcup \left\{ {{\mathbf{v}}_{j{{l}_{j}}}} \right\}$. In the second step, for the remaining subgraphs $\mathcal{G}_{{{{{\ell }'}}_{2}}}^{\left[ 2 \right]}$, $\forall {{{\ell }'}_{2}}\in \left[ 1,{{d}_{2}} \right]$, where $\left( {{R}_{2}}-A_{{{j}'}}^{\left[ 2 \right]} \right)\bigcap P_{{{{{\ell }'}}_{2}}}^{\left[ 2 \right]}=\phi $ for the vertex ${{\mathbf{v}}_{{j}'{{l}_{{{j}'}}}}}\in {{{\eta }'}_{2}}\bigcap P_{{{{{\ell }'}}_{2}}}^{\left[ 2 \right]}$, ${{\mathbf{v}}_{{j}'{{l}_{{{j}'}}}}}$ would be connected to one of the other subgraphs which has completed the vertex connection in the first step. Find a subgraph $\mathcal{G}_{{{\ell }_{2}}}^{\left[ 2 \right]}$ with $\left( {{R}_{2}}-A_{{{j}'}}^{\left[ 2 \right]} \right)\bigcap \mathcal{P}_{{{\ell }_{2}}}^{\left[ 2 \right]}\ne \phi $, and connect the vertex ${{\mathbf{v}}_{{j}'{{l}_{{{j}'}}}}}\in {{{\eta }'}_{2}}\bigcap \mathcal{P}_{{{{{\ell }'}}_{2}}}^{\left[ 2 \right]}$ to one reference vertex ${{\mathbf{v}}_{j{{l}_{j}}}}\in \left( {{R}_{2}}-A_{{{j}'}}^{\left[ 2 \right]} \right)\bigcap \mathcal{P}_{{{\ell }_{2}}}^{\left[ 2 \right]}$, ${{\ell }_{2}}\ne {{{\ell }'}_{2}}$, creating an edge $e_{\left( j{{l}_{j}} \right),\left( {j}'{{l}_{{{j}'}}} \right)}^{\left[ 2 \right]}$ which represents alignment equation of $\operatorname{span}\left( {{\mathbf{H}}_{2j}}{{\mathbf{v}}_{j{{\ell }_{j}}}} \right)=\operatorname{span}( {{\mathbf{H}}_{2{j}'}}{{\mathbf{v}}_{{j}'{{\ell }_{{{j}'}}}}} )$. In this case, the connection operation does not form loops, but merges two subgraphs, $\mathcal{G}_{{{{{\ell }'}}_{2}}}^{\left[ 2 \right]}$ and $\mathcal{G}_{{{\ell }_{2}}}^{\left[ 2 \right]}$, into one, denoted by $\mathcal{G}_{{{\ell }_{2}}}^{\left[ 2 \right]}$. Then, we have $\mathcal{E}_{{{\ell }_{2}}}^{\left[ 2 \right]}=\mathcal{E}_{{{\ell }_{2}}}^{\left[ 2 \right]}\bigcup \mathcal{E}_{{{{{\ell }'}}_{2}}}^{\left[ 2 \right]}\bigcup \left\{ e_{\left( j{{l}_{j}} \right),\left( {j}'{{l}_{{{j}'}}} \right)}^{\left[ 2 \right]} \right\}$. Update $A_{{{j}'}}^{\left[ 2 \right]}$ and ${{Q}^{\left[ 2 \right]}}$ by $A_{{{j}'}}^{\left[ 2 \right]}\leftarrow A_{{{j}'}}^{\left[ 2 \right]}\bigcup \left\{ {{\mathbf{v}}_{j{{l}_{j}}}} \right\}$ and ${{Q}^{\left[ 2 \right]}}\leftarrow {{Q}^{\left[ 2 \right]}}-1$, respectively. In the illustrative example, each subgraph holds a reference vertex. Therefore, the connection between ${{{\eta }'}_{2}}$ and ${{R}_{2}}$ is performed in each single subgraph and thus, only the first step is sufficient. Finally, ${{\mathcal{G}}^{\left[ 2 \right]}}$ has been obtained as in Fig. 7.

\begin{figure}[!t]
\centering
\includegraphics[width=2.0in]{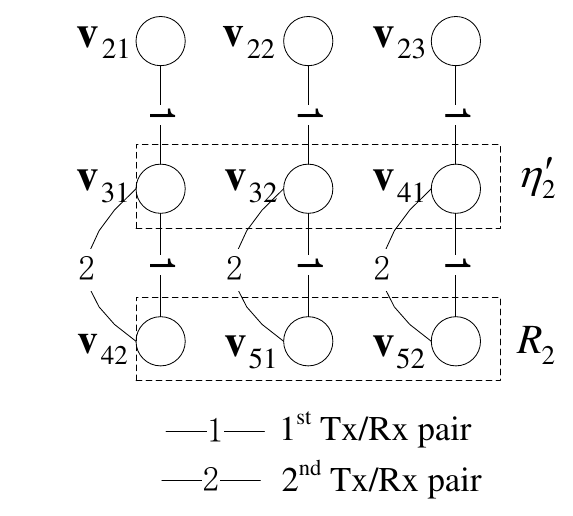}
\vspace*{-10pt}
\caption{Optimal transceiver design of the illustrative example: $\mathcal{G}={{\mathcal{G}}^{\left[ 2 \right]}}$}
\end{figure}

In summary, a total of ${{d}_{2}}$ edges have been added by connecting ${{{\eta }'}_{2}}$ to ${{R}_{2}}$. One edge-addition operation in the first step forms one loop in the corresponding subgraph while the one in the second step merges the corresponding subgraph into another subgraph without forming loops. Assuming that the number of subgraphs subject to the second step is ${{q}^{*}}$, then we have ${{Q}^{\left[ 2 \right]}}={{Q}^{\left[ 1 \right]}}-{{q}^{*}}={{d}_{2}}-{{q}^{*}}$ and each of ${{Q}^{\left[ 2 \right]}}$ subgraphs has one loop. Consequently, we have designed ${{\mathcal{G}}^{\left[ 2 \right]}}$ with only one loop in each subgraph. With $\mathcal{G}={{\mathcal{G}}^{\left[ 2 \right]}}$, the IAC graph $\mathcal{G}$ satisfies Proposition 1 and a set of alignment equations $\Psi $ specified by $\mathcal{G}$ can be correspondingly obtained. Then, the closed-form transceivers, $\left\{ {{\mathbf{V}}_{j}} \right\}_{j=1}^{K}$ and $\left\{ {{\mathbf{U}}_{k}} \right\}_{k=1}^{K}$, can be found as in Section IV-B, while achieving the maximum DoF of $2M=12$ as expected by the theory.

\section{Conclusion}
For $K$-user MIMO interference channel with $M$ Tx/Rx antennas, we have presented a general design principle for a closed-form IAC transceiver with any feasible tuple of DoFs $\left( {{d}_{1}},{{d}_{2}},\cdots ,{{d}_{K}} \right)$. We also have specified the design criterion for ${{k}_{\operatorname{IAC}}}=2$, where the decoded signals of only two receivers are shared through a backhaul link for cancellation, yet achieve the theoretically possible maximum DoFs, which is $2M$.

To our best knowledge, this work is the very first attempt to propose the design that gives the closed-form IAC transceivers for any feasible tuple of DoFs $\left( {{d}_{1}},{{d}_{2}},\cdots ,{{d}_{K}} \right)$. The current design principle can be also applied to IA, as a special case of IAC. As opposed to the existing heuristic algorithms [2-4], which cannot determine the existence of transceiver, i.e., failing to guarantee the performance, our design can achieve a definite DoF for a general system configuration. Moreover, as no iterative steps are involved in the proposed design, it has reduced the computational complexity over heuristic algorithms, turning out to be more practical.  

Considering the optimal design with ${{k}_{\operatorname{IAC}}}=2$, it only requires to send the decoded signals to the first two receivers through a backhaul link for cancellation and thus, the overhead is significantly reduced, while achieving a maximum total DoFs of $2M$. Recall that the maximum total achievable DoFs is limited ${2MK}/{\left( K+1 \right)}\;$ for IA with an equal tuple of DoFs, i.e., ${{d}_{1}}={{d}_{2}}=\cdots ={{d}_{K}}=d$. Furthermore, it is  With a general tuple of DoFs $({{d}_{1}},{{d}_{2}},\cdots ,{{d}_{K}})$, meanwhile, it is available only by a loose upper bound so far, which is given as $\left( 2M-1 \right)$. In other words, we have shown how much performance gain can be achieved by IAC over IA in terms of the total achievable DoF for both equal and general tuples of DoFs, without incurring too much overhead. 

Lastly, we will be evaluating the performance of our proposed IAC design with the imperfect CSI and unreliable backhaul link for a practical system in our future work.

\ifCLASSOPTIONcaptionsoff
  \newpage
\fi

\end{document}